\documentclass[a4paper,11pt]{article}
\pdfoutput=1
\usepackage{jheppub}

\usepackage[utf8]{inputenc}

\allowdisplaybreaks

\usepackage{amssymb}
\usepackage{amsmath}
\usepackage{breqn}
\usepackage{braket}
\usepackage{xcolor}
\usepackage{graphicx}
\usepackage{slashed}

\allowdisplaybreaks

\newcommand{\e}{\epsilon}

\newcommand{\zero}{{(0)}}
\newcommand{\one}{{(1)}}
\newcommand{\two}{{(2)}}
\newcommand{\three}{{(3)}}
\newcommand{\als}{\alpha_s}
\newcommand{\alsmu}{\alpha_s(\mu)}

\newcommand{\Ord}{\mathcal{O}}

\newcommand{\nn}{\nonumber}

\newcommand{\df}{d}

\newcommand{\Gcusp}{\Gamma^{\text{cusp}}}
\newcommand{\Lp}{L_\perp}

\def\cB{\mathcal{B}}
\def\cC{\mathcal{C}}

\def\cI{\mathcal{I}}

\preprint{ZU-TH-46/20}

\title{Unpolarized Quark and Gluon TMD PDFs and FFs at N$^3$LO}

\author[1,2]{Ming-xing Luo,}
\author[3,1]{Tong-Zhi Yang,}
\author[1]{Hua Xing Zhu,}
\author[1]{and Yu Jiao Zhu}

\affiliation[1]{Zhejiang Institute of Modern Physics, Department of Physics, Zhejiang University, Hangzhou, 310027, China}
\affiliation[2]{Complex Systems Division, Beijing Computational Science Research Center, Beijing, 100193, China}
\affiliation[3]{Department of Physics, University of Z\"urich, CH-8057 Z\"urich, Switzerland}

\abstract{In this paper we calculate analytically the perturbative matching coefficients for unpolarized quark and gluon Transverse-Momentum-Dependent~(TMD) Parton Distribution Functions~(PDFs) and Fragmentation Functions~(FFs) through Next-to-Next-to-Next-to-Leading Order~(N$^3$LO) in QCD. The N$^3$LO TMD PDFs are calculated by solving a system of differential equation of Feynman and phase space integrals. The TMD FFs are obtained by analytic continuation from space-like quantities to time-like quantities, taking into account the probability interpretation of TMD PDFs and FFs properly. The coefficient functions for TMD FFs exhibit double logarithmic enhancement at small momentum fraction $z$. We resum such logarithmic terms to the third order in the expansion of $\alpha_s$. Our results constitute important ingredients for precision determination of TMD PDFs and FFs in current and future experiments.}


\begin{document}

\maketitle

\clearpage

\section{Introduction}
\label{sec:introduction}

Understanding the parton structures of hadron is one of the outstanding problem in Quantum ChromodyDynamics~(QCD). TMD PDFs and FFs describe the distribution of parton transverse momentum inside a hadron in parton scattering or decay. Their knowledge is essential to our understanding of the confined motion of parton in nucleons~\cite{Boer:2011fh,Accardi:2012qut,Angeles-Martinez:2015sea}. Thanks to factorization and evolution~\cite{Dokshitzer:1978yd,Parisi:1979se,Collins:1981uw,Collins:1984kg,Ji:2004xq,Ji:2004wu,Bozzi:2005wk,Cherednikov:2007tw,Collins:2011zzd,Becher:2010tm,Echevarria:2012pw,Chiu:2012ir}, TMD PDFs and FFs also enter high precision theoretical prediction for a large variety of observables at  high energy colliders. From pure theoretical point of view, TMD PDFs and FFs are also interesting since they represent light-cone correlation of quantum fields with intrinsic space-like and time-like origin, respectively. Their transparent definition in terms of light-cone correlator also allow higher-order perturbative calculation, from which one can uncover interesting analytic structure of the correlators. 

In the past decade perturbative calculations for TMD distributions have seen rapid development. Next-to-Next-to-Leading Order~(NNLO) corrections to TMD PDFs are first obtained by extraction from expansion of Drell-Yan and Higgs $p_T$ distribution in the relevant kinemtical limit~\cite{Catani:2011kr,Catani:2012qa}. Direct calculation of TMD PDFs and FFs from their light-cone operator definition is difficult due to the existence of unregulated rapidity singularities. Much efforts have been devoted to the  inclusion of rapidity regulator and its associated rapidity factorization~\cite{Collins:1984kg,Ji:2004wu,Collins:2011zzd,Becher:2010tm,Becher:2011dz,Chiu:2012ir,Chiu:2009yx,Echevarria:2015byo,Li:2016axz,Ebert:2018gsn}. Using rapidity regulators, direct computation of TMD PDFs and FFs at NNLO have since become available~\cite{Gehrmann:2012ze,Gehrmann:2014yya,Echevarria:2016scs,Luo:2019hmp,Luo:2019bmw,Gutierrez-Reyes:2019rug}. The progress in NNLO calculation for TMD distributions have allowed many cutting-edge calculations for phenomenology, including fixed-order calculation at NNLO and beyond~\cite{Catani:2007vq,Catani:2009sm,Bozzi:2003jy,Chen:2019lzz,Chen:2019fhs,Cieri:2018oms} and resummation for large logarithms at small $q_T$ at unprecedented N$^3$LL accuracy~\cite{Chen:2018pzu,Bizon:2018foh,Cieri:2018oms,Bizon:2019zgf,Bertone:2019nxa,Bacchetta:2019sam,Ebert:2020dfc,Becher:2020ugp}. We also note that perturbative calculation for TMD quantities suitable for lattice calculation~\cite{Ebert:2019okf,Ji:2019ewn,Ebert:2019tvc,Zhang:2020dbb} has also made important progress recently~\cite{Ebert:2020gxr}. 

Recently a first step towards N$^3$LO TMD distributions have been achieved in \cite{Luo:2019szz} by calculating the unpolarized quark TMD PDFs, extending and significantly improving upon the methods in \cite{Luo:2019hmp,Luo:2019bmw}. Subsequently, both unpolarized quark and gluon TMD PDFs are obtained in \cite{Ebert:2020yqt}, based on an independent method from \cite{Ebert:2020lxs}.
In this paper we continue our calculation in \cite{Luo:2019szz}, and present the N$^3$LO results for unpolarized quark TMD FFs, and unpolarized gluon TMD PDFs and FFs. We also present the results for unpolarized quark TMD PDFs from \cite{Luo:2019szz} for completeness.

Our method for  calculating the  unpolarized gluon TMD PDFs follows \cite{Luo:2019szz}. In order to obtain the results for TMD FFs, we adopt a strategy of analytic continuation proposed in \cite{Chen:2020uvt}. The crucial observation is that although TMD PDFs or FFs are not themselves analytic function of momentum fraction, but the building blocks, space-like and time-like splitting amplitudes, are. At N$^3$LO, there are four distinct contributions to TMD PDFs or FFs, namely the triple real part, the double real-virtual part, the double virtual-real part, and the virtual squared-real part.  Ref.~\cite{Chen:2020uvt} shows that (a) the analytic continuation for the triple real part is trivial since it doesn't involve loop integrals; (b) the analytic continuation for the double real-virtual is also trivial at this order since the continuation of virtual loop only generate $i \pi$ terms which cancel in the sum with complex conjugate. (c) the analytic continuation of double virtual-real part and virtual squared-real part is non-trivial. But they are also simple to calculate since they only involve a one-particle phase space integral. It is suggested in \cite{Chen:2020uvt} that one can calculate these two parts using the corresponding space-like or time-like splitting amplitudes, instead of trying to analytically continue them. Using this approach, Ref.~\cite{Chen:2020uvt} determines the complete time-like splitting functions at NNLO from the space-like counterpart, including the off-diagonal $P_{qg}^{T(2)}$, which cannot be completely fixed with previous methods~\cite{Mitov:2006ic,Moch:2007tx,Almasy:2011eq}. With the complete NNLO time-like splitting functions, Ref.~\cite{Chen:2020uvt} also provides striking evidence for the existence of a generalized Gribov-Lipatov reciprocity relation between space-like and time-like QCD splitting functions in both non-singlet and singlet sector through NNLO. In this paper we use the idea of Ref.~\cite{Chen:2020uvt} to determine not just the splitting functions, but the full time-like TMD FFs through N$^3$LO. Our results for TMD PDFs and FFs are written in terms of familiar harmonic polylogarithms~\cite{Remiddi:1999ew} up to transcendental weight $5$ and allow convenient analytical and numerical manipulation. We provide analytic expression for TMD PDFs and FFs in the threshold limit~($x\,,z \to 1$) and high energy limit~($x\,,z \to 0$). In the high energy limit, TMD FFs exhibit double logarithmic enhancement, instead of single logarithmic enhancement as the in the case of TMD PDFs. A recent discussion of the enhanced double logarithms can be found in \cite{Neill:2020bwv}. We resum the large $\ln z$ terms in TMD FFs to the third order in the expansion of strong coupling, following the method of Vogt~\cite{Vogt:2011jv,Kom:2012hd}. We note that a different method based on celestial BFKL equation has also been developed to resum the time-like small-$z$ logarithms to NNLL in \cite{Neill:2020bwv}.

The structure of this paper is as follows. In Sec.~\ref{sec:defin-quark-gluon} we give the necessary operator definition and renormalization for TMD PDFs and FFs. In Sec.~\ref{sec:n3lo-coeff-unpol} we present the N$^3$LO results for unpolarized quark and gluon TMD PDFs and FFs. Since the expressions are lengthy, we present in the text only a numerical fit to these functions, valid to 0.1 percent in the range of $0 < x\,, z<1$.
The full analytic expressions are provided as ancillary files in the arXiv submission of this paper. We also give the analytic expressions in the threshold limit in this section. In Sec.~\ref{sec:small-x-expansion} we investigate the high energy limit of TMD PDFs and FFs, $x\,, z \to 0$. We provide explicit analytic expressions, showing that TMD FFs are more singular than TMD PDFs from fixed-order point of view, namely double logs in contrast to single logs. We resum the small $z$ logarithms for TMD FFs to the third order in the the expansion of $\alpha_s$. We conclude in Sec.~\ref{sec:conclusion}.

\section{Definition of quark and gluon TMD PDFs and FFs}
\label{sec:defin-quark-gluon}
In this section we give the necessary operator defintion for unpolarized quark and gluon TMD PDFs and FFs. We also give the renormalization counter terms and specify the zero-bin subtraction.

\subsection{Operator definitions for TMD PDFs }
\label{sec:pdf-def}
We begin with the bare TMD PDFs for unpolarized quark and gluon, which can be defined in terms of SCET~\cite{Bauer:2000ew,Bauer:2000yr,Bauer:2001yt,Bauer:2002nz,Beneke:2002ph} collinear fields
\begin{align}
  \label{eq:PDFdef}
   \mathcal{B}_{q/N}^{\rm bare}(x,b_\perp) = &
\int \frac{db^-}{2\pi} \, e^{-i x b^- P^{+}} 
 \langle N(P) | \bar{\chi}_n(0,b^-,b_\perp) \frac{\slashed{\bar{n}}}{2} \chi_n(0) | N(P) \rangle \, ,
  \nn\\
  {\cal B}_{g/N}^{{\rm bare}, \mu \nu}(x,b_\perp) =&
   - x P_+ \int \frac{d b^-}{2 \pi} e^{- i x b^- P^+} \langle N (P) | {\cal A}_{n \perp}^{a,\mu} (0, b^-, b_\perp) {\cal A}_{n \perp}^{a,\nu}(0) | N(P) \rangle \,,
\end{align}
where $N(P)$ is a hadron state with momentum $P^\mu = (\bar{n} \cdot P) n^\mu/2 = P^+ n^\mu/2$, with $n^\mu = (1, 0, 0, 1)$ and $\bar{n}^\mu = (1, 0, 0, -1)$,
 $\chi_n = W_n^\dagger \xi_n$ is the gauge invariant collinear quark field~\cite{Bauer:2001ct} in SCET, 
constructed from collinear quark field $\xi_n$ and path-ordered collinear Wilson line $W_n(x) = {\cal P} \exp \left(i g \int_{-\infty}^0 ds\, \bar{n} \cdot A_n (x + \bar{n} s) \right)$,
and $A_{n\perp}^{a,\mu}$ is the gauge invariant collinear gluon field with color index $a$ and Lorentz index $\mu$. 

For sufficiently small $b_\perp$, the TMD PDFs in Eq.~\eqref{eq:PDFdef} admit operator product expansion onto the usual collinear PDFs,
\begin{align}
  \label{eq:PDFOPE}
    \mathcal{B}_{q/N}^{\rm bare}(x,b_\perp) =& \sum_i  \int_x^1 \frac{d\xi}{\xi} \, \mathcal{I}_{qi}^{\rm bare}(\xi,b_\perp) \, \phi_{i/N}^{\rm bare}(x/\xi) + \text{power corrections} \, ,
    \nn\\
 \mathcal{B}_{g/N}^{{\rm bare}, \mu \nu}(x,b_\perp) =& \sum_i \int_x^1 \frac{d \xi}{\xi} {\cal I}_{gi}^{{\rm bare}, \mu \nu} (\xi, b_\perp) \phi_{i/N}^{\rm bare}(x/\xi) + \text{power corrections}\,,
\end{align}
where the summation is over all parton flavors $i$. 
The perturbative matching coefficients $\mathcal{I}_{qi}^{\rm bare}(\xi,b_\perp)$ and $\cI^{{\rm bare}, \mu \nu}_{gi}(\xi, b_\perp)$ in Eq.~\eqref{eq:PDFOPE} are independent of the actual hadron $N$. 
In practical calculations, one can replace the hadron $N$ with a partonic state $j$. 
Furthermore, the usual bare partonic collinear PDFs are just $\phi^{\rm bare}_{i/j}(x) = \delta_{ij} \delta(1-x)$, therefore
\begin{align}
\cI^{\rm bare}_{qi}(x, b_\perp) = \cB^{\rm bare }_{qi}(x, b_\perp) \,,\quad\quad
\cI^{{\rm bare}, \mu \nu}_{gi}(x, b_\perp) = \cB^{{\rm bare} , \mu \nu}_{gi}(x, b_\perp) \,
\end{align}
up to power correction terms.

For gluon coefficient functions, one can perform a further decomposition into two independent Lorentz structures in $d = 4 - 2 \e$ dimension,
\begin{align}
  \label{eq:PDFdecomposition}
  {\cal I}_{gi}^{{\rm bare}, \mu\nu} (\xi, b_\perp) = \frac{g_\perp^{\mu\nu}}{d-2} 
{\cal I}_{gi}^{{\rm bare}}(\xi, b_T) + \left(\frac{g_\perp^{\mu\nu}}{d-2} + \frac{b_\perp^\mu b_\perp^\nu}{b_T^2} \right) 
{\cal I'}_{gi}^{{\rm bare}}(\xi, b_T) \,,
\end{align}
where we have defined two scalar form factors,
${\cal I}_{gi}^{\rm bare}$, the unpolarized gluon coefficient functions,
 and ${\cal I}^{'\rm bare}_{gi}$, the linearly-polarized gluon coefficient functions. They can be projected out using
\begin{align}
\label{eq:PDFProjection}
\mathcal{I}^{{\rm bare}}_{gi}(\xi, b_T) &= g_\perp^{ \mu \nu} {\cal I}_{gi}^{{\rm bare}, \mu\nu} (\xi, b_\perp)\,, \nonumber \\
\mathcal{I'}^{{\rm bare}}_{gi}(\xi, b_T) &=  \frac{1}{d-3} \left[  g_\perp^{\mu \nu} + \left(   d-2\right) \frac{b_\perp^\mu b_\perp^\nu}{b_T^2} \right] {\cal I}_{gi}^{{\rm bare}, \mu\nu} (\xi, b_\perp) \,, 
\end{align}
with $ b_T^2 = -b_\perp^2 > 0 $ and $b_T = \sqrt{b_T^2} \,.$ 
We focus on unpolarized TMD distributions for the current paper, and the results for the linearly-polarized gluon distribution are left for future work.
\subsection{Operator definitions for TMD FFs }
\label{sec:ff-def}
To specify the definition for gluon TMD FFs, it is necessary to specify a reference frame first. This is in contrast to PDFs, where the incoming hadron provides a canonical reference frame for transverse momentum. In the case of TMD FFs, two different frames can be defined, the hadron frame and the parton frame~\cite{Collins:2011zzd,Luo:2019hmp,Luo:2019bmw}.  In the hadron frame, where the detected hadron has zero transverse momentum, an operator definition for gluon TMD FFs can be written down
\begin{align}
  \label{eq:FF_hadron_Frame}
{\cal D}_{N/q}^{\rm bare} (z, b_\perp)  =& \frac{1}{z}  \sum_{X} \int \frac{db^-}{2\pi}   e^{iP^+  b^-  / z }  
  \langle 0 | 
 \bar \chi_{n}(0,b^-,b_\perp) |  N(P),X \rangle \frac{\slashed{\bar{n}}}{2} \langle  N(P),X | \chi_{n}(0) | 0 \rangle  \,,
 \nn\\
  {\cal D}_{N/g}^{{\rm bare},\mu\nu} (z, b_\perp) = &
- \frac{P_+}{z^2} \sum_X \int \frac{db^-}{2 \pi} 
e^{i P^+  b^- /z} \langle 0 |
{\cal A}_{n\perp}^{a,\mu}(0, b^-, b_\perp) | N(P), X \rangle
\langle N(P), X | {\cal A}_{n\perp}^{a,\nu}(0) | 0 \rangle \,,
\end{align}
where $P^\mu = (\bar n \cdot P) n^\mu/2 =  P^+ n^\mu/2$ is the momenta of the final state detected hadron. 
Again, we focus on unpolarized partonic TMD distributions, the projection to unpolarized gluon distributions as in Eq.~\eqref{eq:PDFdecomposition} is understood from now on. 

In practical calculations, in particular for FF renormalization, it's also convenient to define the fragmentation functions in the parton frame,
 where the parton which initiates the fragmentation has zero transverse momentum.
  The parton frame TMDFFs are related to the hadron frame ones by~\cite{Collins:2011zzd,Luo:2019hmp}
\begin{align}
  \label{eq:partontohadron}
  {\cal F}_{j/i}^{{\rm bare}} (z, b_\perp/z) = z^{2 - 2 \e} {\cal D}_{j/i}^{{\rm bare}} (z, b_\perp)  \,,
\end{align}
where we denote the bare partonic TMD FFs in the parton frame by $ {\cal F}_{j/i}^{{\rm bare}}$. Our N$^3$LO results for TMD FFs will be given in the hadron frame, by choosing the argument of the parton frame coefficient to be $b_\perp/z$. 

\subsection{Renormalization counter terms and zero-bin subtraction}
\label{sec:counter-pdf}
The TMD PDFs or TMD FFs, as well as their matching coefficients, contain both UV and rapidity divergences.
 We adopt dimensional regularization for the UV, and exponentional regularization~\cite{Li:2016axz} for the rapidity divergences. 
 After coupling constant renormalization in $\overline{\rm MS}$ scheme,
 the $\alpha_s$-renormalized matching coefficients still contains overlapping contributions between collinear and soft modes which is removed by a zero-bin subtraction~\cite{Manohar:2006nz}.
 After this, the remaining UV divergences are removed by  multiplicative renormalization counter terms.
 After UV subtraction and zero-bin subtraction described above,
 the TMD PDFs or FFs still contain collinear divergence due to the tagged hadron in initial state or final state,
 and the remaining infrared poles are absorbed into the partonic dimensional regularized collinear PDFs 
 \begin{align}
\phi_{ij}(x, \alpha_s) 
=& \delta_{ij} \delta(1-x) - \frac{\alpha_s}{4 \pi} \frac{P^\zero_{ij}(x)}{\epsilon} 
\nn\\
+&  \left(\frac{\alpha_s}{4 \pi}\right)^2 \bigg[ \frac{1}{2 \epsilon^2} \biggl( \sum_k P^\zero_{ik} \otimes P^\zero_{k j}(x) + \beta_0 P^\zero_{ij} (x)  \biggl) - \frac{1}{2\epsilon} P^\one_{ij}(x)  \bigg]  
\nn \\
+&  \left(\frac{\alpha_s}{4 \pi}\right)^3 \bigg[  \frac{-1}{6 \epsilon^3} \biggl(  \sum_{m \,, k }P^{\zero}_{im} \otimes P^{\zero}_{mk} \otimes P^{\zero}_{k j}(x)  + 3 \beta_0 \sum_k P^{\zero}_{ik} \otimes P^{\zero}_{kj}(x) +2 \beta_0^2 P^{\zero}_{ij}(x) \biggl)  
\nn \\ 
+& \frac{1}{6 \epsilon^2} \biggl( \sum_k P^{\zero}_{ik} \otimes P^{\one}_{kj}(x) 
+ 2 \sum_k P^{\one}_{ik} \otimes P^{\zero}_{kj} (x)  + 2  \beta_0 P^{\one}_{ij}(x) +  2  \beta_1 P^{\zero}_{ij}(x) \biggl)
\nn\\
 -&\frac{1}{3 \epsilon} P^\two_{ij}(x) \bigg]
 +\Ord(\alpha_s^3)  \,,
\end{align}
 or the FFs
\begin{align}
d_{ij}(z, \alpha_s) 
=& \delta_{ij} \delta(1-z) - \frac{\alpha_s}{4 \pi} \frac{P^{T\zero}_{ij}(z)}{\epsilon} 
\nn\\
+&  \left(\frac{\alpha_s}{4 \pi}\right)^2 \bigg[ \frac{1}{2 \epsilon^2} \biggl( \sum_k P^{T\zero}_{ik} \otimes P^{T\zero}_{k j}(z) + \beta_0 P^{T\zero}_{ij} (z)  \biggl) - \frac{1}{2\epsilon} P^{T\one}_{ij}(z)  \bigg]  
\nn \\
+&  \left(\frac{\alpha_s}{4 \pi}\right)^3 \bigg[  \frac{-1}{6 \epsilon^3} \biggl(  \sum_{m \,, k }P^{T\zero}_{im} \otimes P^{T\zero}_{mk} \otimes P^{T\zero}_{k j}(z)  + 3 \beta_0 \sum_k P^{T\zero}_{ik} \otimes P^{T\zero}_{kj}(z) +2 \beta_0^2 P^{T\zero}_{ij}(z) \biggl)  
\nn \\ 
+& \frac{1}{6 \epsilon^2} \biggl( 2\sum_k P^{T\zero}_{ik} \otimes P^{T\one}_{kj}(z) 
+  \sum_k P^{T\one}_{ik} \otimes P^{T\zero}_{kj} (z)  + 2  \beta_0 P^{T\one}_{ij}(z) +  2  \beta_1 P^{T\zero}_{ij}(z) \biggl)
\nn\\
 -&\frac{1}{3 \epsilon} P^{T\two}_{ij}(z) \bigg]
 +\Ord(\alpha_s^3)  \,,
\end{align}
where $P_{ij}^{(n)}$ is the $(n+1)$-loop space-like splitting function~\cite{Moch:2004pa,Vogt:2004mw}, which are also known to the same accuracy in a massive environment~\cite{Ablinger:2014nga,Ablinger:2017tan}.
$P_{ij}^{T (n)}$ is the $(n+1)$-loop time-like splitting function~\cite{Moch:2007tx,Almasy:2011eq,Mitov:2006ic,Chen:2020uvt},
 and the symbol $\otimes$ is denoted as the convolution of two functions
\begin{align}
  f(z,\cdots) \otimes g(z,\cdots) \equiv \int_z^1 \frac{d\xi}{\xi} \, f(\xi,\cdots) \, g(z/\xi,\cdots) \, .
\end{align}
 
 The steps above can be summarized as the following collinear factorization formulas
\begin{align}
  \label{eq:mass-fac-form}
\frac{1}{Z_i^B}  \frac{{{\cal B }}_{i/j}^{{\rm bare}}(x,b_\perp,\mu,\nu)}{\mathcal{S}_{0 \rm b} } = & \sum_k \mathcal{I}_{i k}(x,b_\perp,\mu,\nu) \otimes \phi_{kj}(x,\mu)  \,,
\nn\\
\frac{1}{Z_i^B}  \frac{{{\cal F }}_{j/i}^{{\rm bare}}(z,b_\perp/z,\mu,\nu)}{\mathcal{S}_{0 \rm b} } = & \sum_k  d_{jk}(z,\mu)\otimes \mathcal{C}_{ki} (z, b_\perp/z,\mu,\nu)  \,.
\end{align}
where ${{\cal B }}_{i/j}^{{\rm bare}}$ (${{\cal F }}_{j/i}^{{\rm bare}}$) and $\mathcal{S}_{0 \rm b}(\alpha_s)$ are the bare TMD PDFs~(FFs) and bare zero-bin soft function, 
and $Z_i^B$ (see in Sec.~\ref{sec:RC}) are the multiplicative operator renormalization constants for $i=q,g$.  All the quantities are expressed in terms of the renormalized strong coupling $\alpha_s$.
The zero-bin soft function is the same as TMD soft function, which is known to N$^3$LO in Ref.~\cite{Li:2016ctv}.
Note also that the time-like TMD soft function is identical to the space-like one~\cite{Moult:2018jzp,Zhu:2020ftr} up to this order, so we have a universal soft function up to $\Ord(\alpha_s^3)$.

\section{N$^3$LO coefficients for unpolarized quark and gluon TMDs}
\label{sec:n3lo-coeff-unpol}

In this section we give our results for coefficient functions ${\cal I}_{ij}$ and ${\cal C}_{ij}$. We give only a numeric fit to these functions in the paper, but the full analytic expressions can be found in the ancillary files. 
For TMD FFs, we give the results for ${\cal C}_{ij}$ with an argument $b_\perp/z$, which after divided by $z^2$ are exactly the results in the hadron frame, see Eq.~\eqref{eq:partontohadron}.

\subsection{Renormalization group equations}
\label{sec:RG}
The renormalized coefficient functions obey the following RG equations
\begin{align}
\frac{\df}{\df \ln\mu} \cI_{ji}^{ }(x,b_\perp,\mu,\nu) = 2 \bigg[ \Gcusp_j(\alsmu) \ln\frac{\nu}{xP_{+}} +& \gamma^B_j(\alsmu) \bigg] \cI_{ji}^{}(x,b_\perp,\mu,\nu)
\nn\\
- &2 \sum_k \cI_{jk}^{}(x,b_\perp,\mu,\nu) \otimes P_{ki}(x,\alsmu) \,,
\end{align}
\begin{align}
\frac{\df}{\df \ln\mu} \cC^{ }_{ij}(z,b_\perp/z,\mu,\nu) = 2 \bigg[ \Gcusp_j(\alsmu) \ln\frac{z\nu}{P_{+}} +& \gamma^B_j(\alsmu) \bigg] \cC^{}_{ij}(z,b_\perp/z,\mu, \nu)
\nn\\
-& 2 \sum_k P^T_{ik}(z,\alsmu) \otimes \cC^{}_{kj}(z,b_\perp/z,\mu,\nu) \, .
\label{eq:Imu}
\end{align}
The rapidity evolution equations are~\cite{Chiu:2011qc,Chiu:2012ir}
\begin{align}
\frac{\df}{\df\ln\nu} \cI_{ji}^{}(x,b_\perp,\mu,\nu) =& -2 \left[ \int_{\mu}^{b_0/b_T} \frac{\df\bar{\mu}}{\bar{\mu}} \Gcusp_j(\alpha_s(\bar{\mu})) + \gamma^R_j(\als(b_0/b_T)) \right] \cI_{ji}^{}(x,b_\perp,\mu,\nu) \, ,
\nn\\
\frac{\df}{\df\ln\nu} \cC^{}_{ij}(z,b_\perp/z,\mu,\nu) =& -2 \left[ \int_{\mu}^{b_0/b_T} \frac{\df\bar{\mu}}{\bar{\mu}} \Gcusp_j(\alpha_s(\bar{\mu})) + \gamma^R_j(\als(b_0/b_T)) \right] \cC^{}_{ij}(z,b_\perp/z,\mu, \nu) \,.
\label{eq:Inu}
\end{align}
Expanding the perturbative coefficient functions in terms of $\alpha_s/(4 \pi)$,
 the solution to these evolution equations up to $\Ord(\alpha_s^3)$ reads,
\begin{align}
\cI^\zero_{ji}(x,b_\perp,&\,\mu,\nu) = \delta_{ji} \delta(1-x) \, ,\nn
\\
\cI^\one_{ji}(x,b_\perp,&\,\mu,\nu) = \left( - \frac{\Gcusp_0}{2} \Lp L_Q + \gamma_0^B \Lp + \gamma_0^R L_Q \right) \delta_{ji} \delta(1-x) - P_{ji}^\zero(x) \Lp + I_{ji}^\one(x) \, , \nn
\\
\cI^\two_{ji}(x,b_\perp,&\,\mu,\nu) =  \bigg[ \frac{1}{8} \left( -\Gcusp_0 L_Q + 2\gamma^B_0 \right) \left( -\Gcusp_0 L_Q + 2\gamma^B_0 + 2\beta_0 \right) \Lp^2
\nn\\
+& \left(  (-\Gcusp_0 L_Q + 2\gamma^B_0 + 2\beta_0) \frac{\gamma_0^R}{2} L_Q 
-\frac{\Gcusp_1}{2} L_Q + \gamma^B_1  \right) \Lp 
\nn\\
+& \frac{(\gamma_0^R)^2}{2} L_Q^2 + \gamma_1^R L_Q \bigg] \, \delta_{ji} \delta(1-x) 
+ \bigg( \frac{1}{2} \sum_l P^\zero_{jl} \otimes P^\zero_{li}(x) \nn
\\
+& \frac{P^\zero_{ji}(x)}{2} (\Gcusp_0 L_Q - 2\gamma_0^B - \beta_0) \bigg) \Lp^2 + \bigg[ -P^\one_{ji}(x) - P^\zero_{ji}(x) \gamma_0^R L_Q 
\nn\\
- &\sum_l I^\one_{jl} \otimes P^\zero_{li}(x) \nn
+ \left( -\frac{\Gcusp_0}{2} L_Q + \gamma_0^B + \beta_0 \right) I^\one_{ji}(x) \bigg] \Lp + \gamma_0^R L_Q I^\one_{ji}(x) + I^\two_{ji}(x) \, ,\nn
\\
\cI^\three_{ji}(x,b_\perp,&\,\mu,\nu) = \Lp^3\bigg[
\left(\frac{1}{2}\beta_0+\frac{1}{4}(2 \gamma^B_0-\Gcusp_0 L_Q) \right) \sum_{l}P^\zero_{jl} \otimes P^\zero_{li}(x) 
\nn\\
-&\frac{1}{6} \sum_{l\,k } P^\zero_{jl} \otimes P^\zero_{lk} \otimes P^\zero_{ki}(x) 
+\delta_{ji} \delta(1-x) \left( \frac{1}{6}\beta_0^2(2 \gamma^B_0-\Gcusp_0 L_Q)\right.
\nn\\
+&\left.\frac{1}{8} \beta_0(2 \gamma^B_0-\Gcusp_0 L_Q)^2+\frac{1}{48} (2 \gamma^B_0-\Gcusp_0 L_Q)^3 \right) \nn
\\
+&P^\zero_{ji}  \left( -\frac{1}{2} \beta_0 (2 \gamma^B_0-\Gcusp_0 L_Q)-\frac{1}{3}\beta_0^2-\frac{1}{8} (2 \gamma^B_0-\Gcusp_0 L_Q)^2 \right) \nn
\bigg]
\\
+&\Lp^2\bigg[ 
\left(-\frac{3}{2} \beta_0-\frac{1}{2}(2 \gamma^B_0-\Gcusp_0 L_Q) \right)\sum_l I^\one_{jl} \otimes P^\zero_{li} (x) 
\nn\\
+&\frac{1}{2} \sum_{l\,k } I^\one_{jl} \otimes P^\zero_{lk} \otimes P^\zero_{ki}(x)
+\frac{1}{2} \sum_l P^\zero_{jl} \otimes P^\one_{li}(x)
+\frac{1}{2}\sum_l P^\one_{jl} \otimes P^\zero_{li}(x) 
\nn\\
+&P^\zero_{ji}(x) \left(-\frac{1}{2}\beta_1-\frac{1}{2}(2 \gamma^B_1-\Gcusp_1 L_Q) \right)
+\delta_{ji} \delta(1-x)  \left ( \frac{1}{4} \beta_1(2 \gamma^B_0-\Gcusp_0 L_Q)\right.
\nn\\
+&\left.\frac{1}{2}\beta_0(2 \gamma^B_1-\Gcusp_1 L_Q)+\frac{1}{4}(2 \gamma^B_0-\Gcusp_0 L_Q)(2 \gamma^B_1-\Gcusp_1 L_Q)
 \right)  \nn
\\
+&I^\one_{ji}(x) \left( \frac{3}{4} \beta_0(2 \gamma^B_0-\Gcusp_0 L_Q)+\beta_0^2+\frac{1}{8}(2 \gamma^B_0-\Gcusp_0 L_Q)^2  \right)
\nn\\
+&P^\one_{ji}(x) \left(-\beta_0-\frac{1}{2}(2 \gamma^B_0-\Gcusp_0 L_Q) \right)
\bigg] \nn
\\
+&\Lp \bigg[
-\sum_{l } I^\one_{jl} \otimes P^\one_{li} (x)-\sum_{l } I^\two_{jl} \otimes P^\zero_{li} (x) -P^\zero_{ji}(x) \gamma^R_1 L_Q
-P^\two_{ji}(x) 
\nn\\
+&\delta_{ji} \delta(1-x)  \biggl( 2 \beta_0  \gamma_1^R L_Q 
 +\frac{1}{2}\gamma_1^R (2 \gamma^B_0-\Gcusp_0 L_Q)L_Q+\frac{1}{2} (2\gamma_2^B-\Gcusp_2 L_Q) \biggl )
 \nn\\
 +&I^\one_{ji} (x)\left( \beta_1+\frac{1}{2}(2 \gamma^B_1-\Gcusp_1 L_Q) \right) \nn
+I^\two_{ji} (x)\left(2 \beta_0+\frac{1}{2}(2 \gamma^B_0-\Gcusp_0 L_Q) \right)\bigg]
\nn\\
+&\delta_{ji} \delta(1-x) \gamma^R_2 L_Q+I^\one_{ji}(x) \gamma^R_1 L_Q+I^\three_{ji} (x)\,,
\end{align}
where in $\cI_{ji}^{\three}$ we have used $\gamma_0^R = 0$ to simplify the expression and $I^{(n)}_{ji} (z)$ are the scale-independent coefficient functions. We have defined 
\begin{align}
\label{eq:LdefinitionS}
 L_\perp = \ln \frac{b_T^2 \mu^2}{b_0^2} , \quad  L_Q  = 2 \ln \frac{x \, P_+}{\nu}, \quad L_\nu = \ln \frac{\nu^2}{\mu^2} \,,\quad b_0 =2  e^{- \gamma_E}\,.
\end{align}
Similarly, the solution to the fragmentation coefficient functions are
\begin{align}
\label{eq:RGST}
\cC^\zero_{ji}(z,b_\perp/z,&\,\mu,\nu) =  \delta_{ji} \delta(1-z) \, ,\nn
\\
\cC^\one_{ji}(z,b_\perp/z,&\,\mu,\nu) = \left( - \frac{\Gcusp_0}{2} \Lp L_Q + \gamma_0^B \Lp + \gamma_0^R L_Q \right) \delta_{ji} \delta(1-z) - P_{ji}^{T\zero}(z) \Lp + C_{ji}^\one(z) \, , \nn
\\
\cC^\two_{ji}(z,b_\perp/z,&\,\mu,\nu) =  \bigg[ \frac{1}{8} \left( -\Gcusp_0 L_Q + 2\gamma^B_0 \right) \left( -\Gcusp_0 L_Q + 2\gamma^B_0 + 2\beta_0 \right) \Lp^2
\nn\\
+& \left(  (-\Gcusp_0 L_Q + 2\gamma^B_0 + 2\beta_0) \frac{\gamma_0^R}{2} L_Q 
-\frac{\Gcusp_1}{2} L_Q + \gamma^B_1  \right) \Lp 
\nn\\
+& \frac{(\gamma_0^R)^2}{2} L_Q^2 + \gamma_1^R L_Q \bigg] \, \delta_{ji} \delta(1-z) 
+ \bigg( \frac{1}{2} \sum_l P^{T\zero}_{jl} \otimes P^{T\zero}_{li}(z) \nn
\\
+& \frac{P^{T\zero}_{ji}(z)}{2} (\Gcusp_0 L_Q - 2\gamma_0^B - \beta_0) \bigg) \Lp^2 + \bigg[ -P^{T\one}_{ji}(z) - P^{T\zero}_{ji}(z) \gamma_0^R L_Q 
\nn\\
- &\sum_l P^{T\zero}_{jl} \otimes C^\one_{li}(z) \nn
+ \left( -\frac{\Gcusp_0}{2} L_Q + \gamma_0^B + \beta_0 \right) C^\one_{ji}(z) \bigg] \Lp + \gamma_0^R L_Q C^\one_{ji}(z) + C^\two_{ji}(z) \, ,\nn
\\
\cC^\three_{ji}(z,b_\perp/z,&\,\mu,\nu) = \Lp^3\bigg[
\left(\frac{1}{2}\beta_0+\frac{1}{4}(2 \gamma^B_0-\Gcusp_0 L_Q) \right) \sum_{l}P^{T\zero}_{jl} \otimes P^{T\zero}_{li}(z) 
\nn\\
-&\frac{1}{6} \sum_{l\,k } P^{T\zero}_{jl} \otimes P^{T\zero}_{lk} \otimes P^{T\zero}_{ki}(z) 
+\delta_{ji} \delta(1-z) \left( \frac{1}{6}\beta_0^2(2 \gamma^B_0-\Gcusp_0 L_Q)\right.
\nn\\
+&\left.\frac{1}{8} \beta_0(2 \gamma^B_0-\Gcusp_0 L_Q)^2+\frac{1}{48} (2 \gamma^B_0-\Gcusp_0 L_Q)^3 \right) \nn
\\
+&P^{T\zero}_{ji}  \left( -\frac{1}{2} \beta_0 (2 \gamma^B_0-\Gcusp_0 L_Q)-\frac{1}{3}\beta_0^2-\frac{1}{8} (2 \gamma^B_0-\Gcusp_0 L_Q)^2 \right) \nn
\bigg]
\\
+&\Lp^2\bigg[ 
\left(-\frac{3}{2} \beta_0-\frac{1}{2}(2 \gamma^B_0-\Gcusp_0 L_Q) \right)\sum_l P^{T\zero}_{jl} \otimes C^\one_{li} (z) 
\nn\\
+&\frac{1}{2} \sum_{l\,k } P^{T\zero}_{jl} \otimes P^{T\zero}_{lk} \otimes C^\one_{ki}(z)
+\frac{1}{2} \sum_l P^{T\zero}_{jl} \otimes P^{T\one}_{li}(z)
+\frac{1}{2}\sum_l P^{T\one}_{jl} \otimes P^{T\zero}_{li}(z) 
\nn\\
+&P^{T\zero}_{ji}(z) \left(-\frac{1}{2}\beta_1-\frac{1}{2}(2 \gamma^B_1-\Gcusp_1 L_Q) \right)
+\delta_{ji} \delta(1-z)  \left ( \frac{1}{4} \beta_1(2 \gamma^B_0-\Gcusp_0 L_Q)\right.
\nn\\
+&\left.\frac{1}{2}\beta_0(2 \gamma^B_1-\Gcusp_1 L_Q)+\frac{1}{4}(2 \gamma^B_0-\Gcusp_0 L_Q)(2 \gamma^B_1-\Gcusp_1 L_Q)
 \right)  \nn
\\
+&C^\one_{ji}(z) \left( \frac{3}{4} \beta_0(2 \gamma^B_0-\Gcusp_0 L_Q)+\beta_0^2+\frac{1}{8}(2 \gamma^B_0-\Gcusp_0 L_Q)^2  \right)
\nn\\
+&P^{T\one}_{ji}(z) \left(-\beta_0-\frac{1}{2}(2 \gamma^B_0-\Gcusp_0 L_Q) \right)
\bigg] \nn
\\
+&\Lp \bigg[
-\sum_{l } P^{T\one}_{jl} \otimes C^\one_{li} (z)-\sum_{l } P^{T\zero}_{jl} \otimes C^\two_{li} (z) -P^{T\zero}_{ji}(z) \gamma^R_1 L_Q
-P^{T\two}_{ji}(z) 
\nn\\
+&\delta_{ji} \delta(1-z)  \biggl( 2 \beta_0  \gamma_1^R L_Q 
 +\frac{1}{2}\gamma_1^R (2 \gamma^B_0-\Gcusp_0 L_Q)L_Q+\frac{1}{2} (2\gamma_2^B-\Gcusp_2 L_Q) \biggl )
 \nn\\
 +&C^\one_{ji} (z)\left( \beta_1+\frac{1}{2}(2 \gamma^B_1-\Gcusp_1 L_Q) \right) \nn
+C^\two_{ji} (z)\left(2 \beta_0+\frac{1}{2}(2 \gamma^B_0-\Gcusp_0 L_Q) \right)\bigg]
\nn\\
+&\delta_{ji} \delta(1-z) \gamma^R_2 L_Q+C^\one_{ji}(z) \gamma^R_1 L_Q+C^\three_{ji} (z)\,.
\end{align}
We stress again that due to the chosen argument, the expressions given above are for TMD FFs in the hadron frame.
The anomalous dimensions appeared above are identical to those in space-like case and we have  suppressed their dependence on the exact flavor.
The logarithms appeared in the fragmentation coefficient functions are defined as
\begin{align}
\label{eq:LdefinitionT}
 L_\perp = \ln \frac{b_T^2 \mu^2}{b_0^2} , \quad L_Q = 2 \ln \frac{P_+}{ z \, \nu}, \quad L_\nu = \ln \frac{\nu^2}{\mu^2} \,,\quad b_0 =2  e^{- \gamma_E}\,,
\end{align}
which differ from thsoe in Eq.~\eqref{eq:LdefinitionS} only in $L_Q$.
Both space-like and time-like coefficient functions depend on the rapidity regulator being used. 
Rapidity-regulator-independent TMD PDFs and TMD FFs can be obtained by 
multiplying the coefficient functions with the squared root of the TMD soft functions ${\cal S}(b_\perp, \mu, \nu)$~\cite{Luo:2019hmp,Luo:2019bmw}
\begin{align}
  \label{eq:TMD-PDF-FF}
  f_{\perp,ij}(x, b_\perp, \mu) =& {\cal I}_{ij}(x, b_\perp, \mu, \nu) \sqrt{{\cal S}(b_\perp, \mu, \nu)} \,,
\nn\\
 g_{\perp,ij}(z, b_\perp/z,  \mu) = &{\cal C}_{ij}(z, b_\perp/z,  \mu, \nu) \sqrt{{\cal S}(b_\perp, \mu, \nu)} \,.
\end{align}

\subsection{Numerical fits of the N3LO coefficients}
\label{sec:num-fit}
The analytic expressions for the  coefficient functions  will be provided in the ancillary files along with the arXiv submission.
In this section we will present their numerical fits.

The coefficient functions develop end-point divergences both in the threshold and high energy limit.
We first present here the results for leading threshold limit. The results for high energy limit will be discussed in next section. In the $z \to 1$ limit, we have
\begin{equation}
  \label{eq:largez}
 \lim_{z\to1}  \mathcal{I}^{(2)}_{ij}(z)= \lim_{z\to1} \mathcal{C}^{(2)}_{ji}(z) = \frac{2 \gamma^R_{1,i}}{(1-z)_+}\delta_{ij} \,, \quad 
  \lim_{z\to1}  \mathcal{I}^{(3)}_{ij}(z)=\lim_{z\to1} \mathcal{C}^{(3)}_{ji}(z) = \frac{2 \gamma^R_{2,i}}{(1-z)_+} \delta_{ij} \,,
\end{equation}
where $\gamma^R_{1(2)}$ are the two(three)-loop rapidity anomalous dimensions~\cite{Li:2016ctv,Vladimirov:2016dll}. The relation between threshold limit and rapidity anomalous dimension has been anticipated in \cite{Echevarria:2016scs,Lustermans:2016nvk,Billis:2019vxg}.
The explicit expressions up to three-loop read~\cite{Luo:2019szz}
 \begin{align}
\mathcal{I}^{(1)}_{qq}(z)= &\mathcal{C}^{(1)}_{qq}(z)=0\,,
 \nn\\
\mathcal{I}^{(2)}_{qq}(z)=&\mathcal{C}^{(2)}_{qq}(z)=  \frac{1}{(1-z)_+} \bigg[ \left(28 \zeta _3-\frac{808}{27}\right) C_A C_F+\frac{224}{27} C_F N_f T_F \bigg] \,,
\nn \\
\mathcal{I}^{(3)}_{qq}(z)=&\mathcal{C}^{(3)}_{qq}(z)=  \frac{1}{(1-z)_+} \bigg[ 
   \biggl(-\frac{1648 \zeta _2}{81}-\frac{1808 \zeta _3}{27}+\frac{40 \zeta
   _4}{3} 
   +\frac{125252}{729}\biggl) C_A C_F N_f T_F\nonumber 
    \\
   +&\biggl(-\frac{176}{3} \zeta _3 \zeta
   _2+\frac{6392 \zeta _2}{81}  
 +\frac{12328 \zeta _3}{27}+\frac{154 \zeta _4}{3}-192 \zeta
   _5-\frac{297029}{729}\biggl) C_A^2 C_F \nonumber 
   \\
   +& \left(-\frac{608 \zeta _3}{9}-32 \zeta
   _4+\frac{3422}{27}\right) C_F^2 N_f T_F
  + \left(-\frac{128}{9} \zeta
   _3-\frac{7424}{729}\right) C_F N_f^2 T_F^2  
     \bigg] \,.
 \end{align}
 We also found that threshold limit exhibits Casimir scaling up to three loops~($n=1,2,3$),
\begin{align}
\lim_{z\to1}\frac{\mathcal{I}^{(n)}_{gg}(z)}{\mathcal{I}^{(n)}_{qq}(z)}=\lim_{z\to1}\frac{\mathcal{C}^{(n)}_{gg}(z)}{\mathcal{C}^{(n)}_{qq}(z)}=\frac{C_A}{C_F} \,.
\end{align}

\subsubsection{Numerical fit for TMD PDFs}
The analytic expressions of three-loop coefficient functions contain harmonic polylogarithms up to transcendental weight $5$. To facilitate straightforward numerical implementation, we provide a numerical fitting to all the coefficient functions. Following Ref.~\cite{Moch:2017uml}, we use the following elementary functions to fit the results,
\begin{align}
\label{eq:Lzdefinition}
L_x \equiv \ln x\,,\,L_{\bar{x}} \equiv \ln (1-x )\,, \, \bar{x}\equiv 1-x \,. 
\end{align}
For two loop and three loop coefficient functions, we fit the exact results in the region $10^{-6} < x <1-10^{-6}$ (Numerical evaluation of HPLs are made with the \texttt{HPL} package~\cite{Maitre:2005uu}), and we have set the color factor to numerical values in QCD, i.e. 
\begin{align}
C_F = \frac{4}{3}\,,\qquad  C_A = 3\,, \qquad T_F = \frac{1}{2} \,. 
\end{align}
In more detail, we subtract the $x \to 0$ and $x \to 1$ limits up to next-to-next-to-leading power ($x^1 $ and $(1-x)^1$). Then we fit the remaining terms in the region $10^{-6} < x <1-10^{-6}$. Combining the two parts, the fitted results can achieve an accuracy better than $10^{-3}$ for $0<x<1$. We show below the numerical fitting with six significant digits. The full numerical fitting is attached as ancillary files with the arXiv submission.
The one loop scale independent coefficient functions are given by
\begin{subequations}
\begin{align}
I^{(1)}_{qq}(x) =&2.66667 \bar{x}   \,, \\
I^{(1)}_{qg}(x) =& 2 x \, \bar{x}  \,, \\
I^{(1)}_{gq}(x) =& 2.66667 x \,, \\
I^{(1)}_{gg}(x) =&  0 \,, 
\end{align}
\end{subequations}
The two-loop scale independent coefficient function are given by
\begin{dgroup}
\begin{dmath}
I^{(2)}_{qq'}(x) =  \frac{2.64517}{x} -4.56035\, +x^3 \left(-0.0170296 L_x^3-0.143469 L_x^2+1.21562
   L_x-3.60403\right) +x^2 \big(0.000306717 L_x^3-1.77306 L_x^2   
   +5.65477
   L_x+3.57046\big)  +x \left(0.444444 L_x^3 -0.666667 L_x^2-5.33333
   L_x+1.89369\right)+0.444444 L_x^3  
   -0.666667 L_x^2+2.66667
   L_x -0.00131644 x^5+0.0563783 x^4+1.33333 \bar{x}\,,
\end{dmath}  
\begin{dmath}
  I^{(2)}_{q\bar{q}}(x) =  I^{(2)}_{qq'}(x)+ x^3 \left(23.8756 L_x^3-68.5281 L_x^2+391.31 L_x-479.112\right)+x^2
   \big(1.73989 L_x^3+29.5744 L_x^2+207.751 L_x 
   +533.913\big)+x
   \left(-0.148148 L_x^3-0.888889 L_x^2-1.33333
   L_x+6.98894\right)+0.148148 L_x^3-1.33333 L_x
   +1.66959 x^5-60.5553
   x^4-0.444444 \bar{x}-2.90423 \,, 
\end{dmath}
\begin{dmath}   
  I^{(2)}_{qq}(x) =  I^{(2)}_{qq'}(x)+  \left(5.53086 N_f+14.9267\right) \frac{1}{(\bar{x})}_+ +N_f
   \bigg\{x^3 \left(-0.0532042 L_x^2+1.92031 L_x-4.39249\right) 
   +x^2
   \left(0.939547 L_x^2+3.50359 L_x+1.65854\right)+x \left(0.444444
   L_x^2+1.48148 L_x+2.36991\right)+0.444444 L_x^2
   +1.48148
   L_x+0.0178399 x^5-0.246399 x^4+4.24691 \bar{x}-7.90123 \bigg\}+x^3
   \big(3.49597 L_x^3-18.6432 L_x^2+60.163 L_x
   -48.6244\big)+x^2
   \left(-2.49636 L_x^3-11.0306 L_x^2-7.51243
   L_x+59.7912\right)+\left(\bar{x}\right)^3 \left(3.19726 L_{\bar{x}}-1.32635
   L_{\bar{x}}^2\right)
   -7.11111 L_{\bar{x}}^2+\left(\bar{x}\right)^2 \left(13.4628
   L_{\bar{x}}-2.37726 L_{\bar{x}}^2\right)+22.2222 L_{\bar{x}}+\bar{x} \left(3.55556
   L_{\bar{x}}^2-17.7778 L_{\bar{x}}-0.105144\right)
   +x \left(-0.740741 L_x^3-10.
   L_x^2-11.5556 L_x+1.87655\right)-0.740741 L_x^3-2. L_x^2-8.
   L_x
   +0.070919 x^5-2.2589 x^4-10.2974 \,, 
\end{dmath}
\begin{dmath}   
   I^{(2)}_{qg}(x) =  -52.3982\, +0.555556 L_{\bar{x}}^3+\left(\bar{x}\right)^3
   \left(-2.91634 L_{\bar{x}}^3+2.52056 L_{\bar{x}}^2-54.8176
   L_{\bar{x}}\right)
   +\left(\bar{x}\right)^2 \left(0.982654 L_{\bar{x}}^3+2.72223
   L_{\bar{x}}^2-15.0644 L_{\bar{x}}\right)-1.66667 L_{\bar{x}}+\bar{x} \big(-1.11111
   L_{\bar{x}}^3-4.66667 L_{\bar{x}}^2+5. L_{\bar{x}}
   +58.9092\big)+x \left(2.44444
   L_x^3+11.6667 L_x^2+6.66667 L_x-53.6197\right)+0.777778
   L_x^3-1.16667 L_x^2
   +11.3333 L_x+4.0827 x^5-16.7693
   x^4+\frac{5.95164}{x}+x^3 \left(-0.5403 L_x^3+15.2935 L_x^2+21.5425
   L_x-103.137\right) +x^2 \big(-1.00863 L_x^3-20.4656 L_x^2
   -24.7319
   L_x+200.419\big) \,,
\end{dmath}
\begin{dmath}
I^{(2)}_{gq}(x) = N_f \bigg\{-25.4815\, +\bar{x}^3 \left(0.319337 L_{\bar{x}}^2+3.6769 L_{\bar{x}}\right)+\bar{x}^2 \left(1.83746 L_{\bar{x}}^2+6.59938 L_{\bar{x}}\right)+0.888889 L_{\bar{x}}^2+1.18519 L_{\bar{x}}+\bar{x} \left(0.888889 L_{\bar{x}}^2+4.74074 L_{\bar{x}}+8.49383\right)-0.0201765 x^6-0.0606673 x^5-0.710953 x^4+10.7761 x^3-25.0013 x^2+32.0047 x+\frac{11.0617}{x}\bigg\}+25.1431\, +x^3 \left(2.72767 L_x^3+11.0188 L_x^2+67.724 L_x+124.821\right)+x^2 \left(0.001443 L_x^3+0.195266 L_x^2+2.68862 L_x-86.22\right)+x \left(-3.25926 L_x^3+9.33333 L_x^2-39.1111 L_x+48.9377\right)-4.14815 L_x^3+12.4444 L_x^2-66.6667 L_x+\bar{x}^3 \left(8.67537 L_{\bar{x}}^3-26.9743 L_{\bar{x}}^2+101.276 L_{\bar{x}}\right)+\bar{x}^2 \left(-2.72739 L_{\bar{x}}^3-25.4335 L_{\bar{x}}^2-43.8567 L_{\bar{x}}\right)-1.48148 L_{\bar{x}}^3-4.88889 L_{\bar{x}}^2-20.4444 L_{\bar{x}}+\bar{x} \left(-1.48148 L_{\bar{x}}^3-21.7778 L_{\bar{x}}^2-41.7778 L_{\bar{x}}-68.9101\right)-1.2442 x^6+7.84991 x^5-72.9318 x^4-\frac{44.3476}{x} \,,
\end{dmath}

\begin{dmath}
I^{(2)}_{gg}(x) = N_f \bigg\{12.4444 \frac{1}{(\bar{x})}_+ +0.888889 L_x^3+6. L_x^2+24.6667 L_x+x \left(0.888889 L_x^3+3.33333 L_x^2+22.6667 L_x+36.\right)-2. L_{\bar{x}}+\bar{x} \left(2. L_{\bar{x}}+54.2222\right)-29.1111 x^2+\frac{25.1111}{x}-48.4444\bigg\}+33.585 \frac{1}{(\bar{x})}_+-85.3465\, +x^3 \left(-51.5131 L_x^3+433.575 L_x^2-477.56 L_x+2429.\right)+x^2 \left(-4.94833 L_x^3-97.5448 L_x^2-706.464 L_x-1749.4\right)+x \left(-24. L_x^3-33. L_x^2-269. L_x+221.873\right)-12. L_x^3+3. L_x^2-293. L_x+\bar{x}^3 \left(4.98201 L_{\bar{x}}-3.11742 L_{\bar{x}}^2\right)+\bar{x}^2 \left(-67.425 L_{\bar{x}}^2-16.306 L_{\bar{x}}\right)-36. L_{\bar{x}}^2+6. L_{\bar{x}}+\bar{x} \left(18. L_{\bar{x}}^2-6. L_{\bar{x}}-352.048\right)-23.039 x^6+160.219 x^5-756.676 x^4-\frac{99.7822}{x} \,.
\end{dmath}
\end{dgroup}
To present the three-loop scale independent coefficient functions, we first perform the following decompositions,
\begin{align}
 I^{(3)}_{qq'} (x) &= I^*_{qq'}(x) + \frac{d^{ABC} d_{ABC} \, }{32 N_c } I_{d33}(x),   \nonumber \\
 I^{(3)}_{q\bar{q}'}(x) &= I^*_{qq'}(x) - \frac{d^{ABC} d_{ABC} \, }{32 N_c} I_{d33}(x),   \nonumber \\  
I^{(3)}_{qq}(x) &=  I^*_{qq}(x) + I^{(3)}_{qq'}(x) ,   \nonumber  \\
I^{(3)}_{q\bar{q}}(x) &=   C_F \left(C_A-2 C_F\right) I^*_{q\bar{q}}(x) +I^{(3)}_{q\bar{q}'}(x) \,,
\label{eq:Idecomp}
\end{align}
where
\begin{equation}
d^{ABC} d_{ABC} = 4 \mathrm{Tr}[T^A \{T^B , T^C \} ] \mathrm{Tr}[T^A \{ T^B,  T^C\}] = \frac{(N_c^2-1)(N_c^2-4)}{N_c} = \frac{40}{3} \,.
\end{equation} 
The numerical fitting of different color structures are given by 
\begin{dgroup}
\begin{dmath} 
 I^*_{qq'}(x)\, = N_f \bigg\{-50.3634\, +x^3 \left(0.117123 L_x^4-0.314883 L_x^3+3.53761
   L_x^2-10.541 L_x+18.5347\right)+x^2 \big(-0.0026012 L_x^4 
   +0.703183
   L_x^3+0.199736 L_x^2+5.14781 L_x-24.9854\big)+\left(\bar{x}\right)^3
   \left(-0.227266 L_{\bar{x}}^3-0.0305807 L_{\bar{x}}^2-3.18523
   L_{\bar{x}}\right) 
   +\left(\bar{x}\right)^2 \left(-0.000912305 L_{\bar{x}}^3+0.865032
   L_{\bar{x}}^2+1.85446 L_{\bar{x}}\right)+\bar{x} \big(-0.0987654 L_{\bar{x}}^3-0.493827
   L_{\bar{x}}^2-4.21399 L_{\bar{x}} 
   -8.73525\big)+x \big(-0.246914
   L_x^4-1.61317 L_x^3-8.77153 L_x^2-12.0035
   L_x+51.7382\big)-0.246914 L_x^4-1.61317 L_x^3 
   -10.5493
   L_x^2-30.9665 L_x+0.102137 x^5-0.909082
   x^4+\frac{5.88282}{x}\bigg\}+307.912\, +x^3 \big(1.80474
   L_x^5+1.58169 L_x^4  
   +92.844 L_x^3+24.4101 L_x^2+724.086
   L_x+168.701\big)+x^2 \big(-0.00854607 L_x^5-4.88703 L_x^4-33.0209
   L_x^3 
   +188.744 L_x^2-49.3191 L_x+147.078\big)+\left(\bar{x}\right)^3
   \left(-2.52551 L_{\bar{x}}^4+13.1597 L_{\bar{x}}^3-119.84 L_{\bar{x}}^2+333.889
   L_{\bar{x}}\right) 
   +\left(\bar{x}\right)^2 \left(0.0720812 L_{\bar{x}}^4+1.43763
   L_{\bar{x}}^3+24.7149 L_{\bar{x}}^2+192.782 L_{\bar{x}}\right)+\bar{x} \big(0.246914
   L_{\bar{x}}^4+0.54321 L_{\bar{x}}^3+6.03113 L_{\bar{x}}^2 
   +38.031
   L_{\bar{x}}+123.709\big)+x \left(1.27407 L_x^5+2.34568 L_x^4+4.50092
   L_x^3-70.7153 L_x^2-270.973 L_x+167.376\right) 
   -0.592593
   L_x^5+6.79012 L_x^4-46.4127 L_x^3+86.421 L_x^2-470.887
   L_x 
   +\frac{-78.9847 L_x-466.384}{x}+10.7916 x^5-335.475 x^4  \,,  
\end{dmath}

\begin{dmath}
I_{d33}(x)\, = 32. \bigg\{  -x^3 \left(-19156.8 L_x^5+199731.
   L_x^4-2.18614\times 10^6 L_x^3+1.2083\times 10^7
   L_x^2-4.5734\times 10^7 L_x+7.52147\times
   10^7\right)  
    - \left(\bar{x}\right)^3 \left(1023.79
   L_{\bar{x}}^2+121.796 L_{\bar{x}}\right)- \left(\bar{x}\right)^2
   \left(-93.2504 L_{\bar{x}}^2-876.881 L_{\bar{x}}\right) 
    - \bar{x}
   \left(0.0241557 L_{\bar{x}}^2-0.278902
   L_{\bar{x}}-0.582336\right)- x \left(-1.64493
   L_x^3+3.75242 L_x^2+0.397826
   L_x-774.514\right)  
    -0.0333333 L_x^5+0.0833333
   L_x^4-0.132844 L_x^3-0.339443 L_x^2-12.6418
   L_x+8847.28 x^5-729429. x^4-13.0667  - x^2 \left(-538.817 L_x^5-25014.2
   L_x^4-500678. L_x^3-5.37744\times 10^6
   L_x^2-3.09137\times 10^7 L_x-7.59345\times
   10^7\right)   \bigg\}  \,, 
\end{dmath}

\begin{dmath}   
I^*_{qq}(x) \,= \left(-9.09324 N_f^2+154.257 N_f+140.136\right)
 \frac{1}{(\bar{x})}_+  +N_f^2 \bigg\{x^3 \left(0.666009
   L_x^3-4.08064 L_x^2+13.5087 L_x-13.4503\right) 
   +x^2 \left(-0.610414
   L_x^3-2.30736 L_x^2+0.622374 L_x+20.7035\right)+x \big(-0.329218
   L_x^3-0.855967 L_x^2-1.18519 L_x  
   -6.71567\big)-0.329218
   L_x^3-2.43621 L_x^2-5.66255 L_x-0.000436414 x^5-0.405417 x^4-14.1598
   \bar{x}+15.8093\bigg\} 
   +N_f \bigg\{x^3 \left(-11.0536 L_x^4+57.1168
   L_x^3-406.207 L_x^2+1183.43 L_x-2303.63\right)+x^2 \big(3.38122
   L_x^4+37.927 L_x^3 
    +212.612 L_x^2+982.488 L_x+2012.66\big)+2.107
   L_{\bar{x}}^3+\left(\bar{x}\right)^3 \left(0.189723 L_{\bar{x}}^3+1.80938
   L_{\bar{x}}^2-2.62339 L_{\bar{x}}\right) 
   +10.3704 L_{\bar{x}}^2+\left(\bar{x}\right)^2
   \left(0.69507 L_{\bar{x}}^3+5.12831 L_{\bar{x}}^2-15.0284
   L_{\bar{x}}\right)-10.4458 L_{\bar{x}}+\bar{x} \big(-1.0535 L_{\bar{x}}^3-7.60494
   L_{\bar{x}}^2 
   +48.0993 L_{\bar{x}}+439.611\big)+x \left(0.855967
   L_x^4+12.6639 L_x^3+16.4326 L_x^2+43.9956
   L_x+164.293\right)+0.855967 L_x^4 
   +11.9396 L_x^3+59.1809
   L_x^2+187.752 L_x-1.34066 x^5+82.5779 x^4-312.745\bigg\}+x^3
   \big(15.3681 L_x^5-12.7492 L_x^4 
   +195.91 L_x^3+1220.54 L_x^2-4589.86
   L_x+14892.1\big)+x^2 \big(-2.96846 L_x^5-62.4209 L_x^4-385.992
   L_x^3-2295.84 L_x^2 
   -8061.05 L_x-14175.7\big)-34.7654
   L_{\bar{x}}^3+\left(\bar{x}\right)^3 \left(-9.48494 L_{\bar{x}}^3-52.1748
   L_{\bar{x}}^2+111.502 L_{\bar{x}}\right)-5.09037 L_{\bar{x}}^2 
   +\left(\bar{x}\right)^2
   \left(-11.021 L_{\bar{x}}^3-85.3807 L_{\bar{x}}^2+413.664
   L_{\bar{x}}\right)+637.843 L_{\bar{x}}+\bar{x} \big(-7.90123 L_{\bar{x}}^3+9.5085
   L_{\bar{x}}^2-487.943 L_{\bar{x}} 
   -2438.69\big)+x \left(-0.301235
   L_x^5-13.037 L_x^4-60.4295 L_x^3+90.2398 L_x^2+400.357
   L_x+265.017\right)-0.301235 L_x^5 
   -8.69136 L_x^4-69.1867
   L_x^3-286.991 L_x^2-913.865 L_x+9.73661 x^5-868.539 x^4+1033.69 \,,
\end{dmath}

\begin{dmath}  
 I^*_{q\bar{q}}(x) \, = \frac{9 N_f}{4} \bigg\{x^3 \left(-37.3313 L_x^4+69.5328 L_x^3-1189.55 L_x^2+2193.6
   L_x-4962.46\right)+x^2 \big(-0.0538545 L_x^4+8.68065 L_x^3 
   +179.886
   L_x^2+1428.28 L_x+4299.55\big)+0.101532 \left(\bar{x}\right)^3
   L_{\bar{x}}+0.0172219 \left(\bar{x}\right)^2 L_{\bar{x}}+\bar{x} \left(0.395062
   L_{\bar{x}}+0.460905\right) 
   +x \left(0.131687 L_x^4+0.877915 L_x^3+4.86623
   L_x^2+11.488 L_x-9.51143\right)-0.131687 L_x^4-0.943759
   L_x^3-3.41767 L_x^2 
   -4.49137 L_x-16.6625 x^5+688.059
   x^4+1.02498\bigg\}+ \frac{9}{4} \bigg\{ x^3 \big(-1385.38 L_x^5+15741.4 L_x^4-165914.
   L_x^3+944859. L_x^2 
   -3.53957\times 10^6 L_x+5.88567\times
   10^6\big)+x^2 \big(-42.7578 L_x^5-1977.89 L_x^4-39497.9
   L_x^3-422824. L_x^2 
   -2.42091\times 10^6 L_x-5.92341\times
   10^6\big)+3.3635 \left(\bar{x}\right)^3 L_{\bar{x}}-1.78893
   \left(\bar{x}\right)^2 L_{\bar{x}}+\bar{x} \left(-4.98979 L_{\bar{x}}-13.046\right) 
   +x
   \left(0.0148148 L_x^5-1.48148 L_x^4-13.6105 L_x^3-83.3483
   L_x^2-64.0087 L_x+105.151\right)-0.0148148 L_x^5 
   +1.08642
   L_x^4+10.911 L_x^3+17.5783 L_x^2+6.83186 L_x-246.269 x^5+37870.3
   x^4+15.737 \bigg\} \,,
\end{dmath}   
   
\begin{dmath}      
 I^{(3)}_{qg}(x)\, = N_f \bigg\{532.389\, +x^3 \left(3.23607 L_x^5-4.68217 L_x^4+120.483
   L_x^3-119.874 L_x^2+453.908 L_x+442.25\right) 
    +x^2 \left(-0.0565211
   L_x^5-3.10462 L_x^4-20.1426 L_x^3-202.066 L_x^2-561.287
   L_x+424.055\right)-0.154321 L_{\bar{x}}^4 
   -0.823045
   L_{\bar{x}}^3+\left(\bar{x}\right)^3 \left(2.61221 L_{\bar{x}}^4-9.9784
   L_{\bar{x}}^3+116.968 L_{\bar{x}}^2-267.94 L_{\bar{x}}\right)+1.48131
   L_{\bar{x}}^2 
   +\left(\bar{x}\right)^2 \left(-0.378008 L_{\bar{x}}^4-4.9419
   L_{\bar{x}}^3-38.9532 L_{\bar{x}}^2-203.044 L_{\bar{x}}\right)+22.2518 L_{\bar{x}}+\bar{x}
   \big(0.308642 L_{\bar{x}}^4+2.38683 L_{\bar{x}}^3 
   +3.85219 L_{\bar{x}}^2-18.9149
   L_{\bar{x}}-275.431\big)+x \big(-0.355556 L_x^5-3.65432 L_x^4-21.6379
   L_x^3-75.7833 L_x^2-71.9808 L_x  
   -1040.7\big)+0.177778 L_x^5+1.2716
   L_x^4+14.2634 L_x^3+67.2339 L_x^2+216.898 L_x-2.00607 x^5-287.428
   x^4+\frac{11.9546}{x}\bigg\} 
   -3636.55\, +x^3 \left(-259.099
   L_x^5+3952.96 L_x^4-34085. L_x^3+221224. L_x^2-768027.
   L_x+594756.\right) 
   +x^2 \left(-10.1315 L_x^5-463.996 L_x^4-8964.71
   L_x^3-91774.6 L_x^2-536653. L_x+96171.9\right)-0.925926
   L_{\bar{x}}^5 
   +2.36111 L_{\bar{x}}^4+14.151 L_{\bar{x}}^3+\left(\bar{x}\right)^3
   \left(-264.815 L_{\bar{x}}^5+1791.21 L_{\bar{x}}^4-23536.1 L_{\bar{x}}^3+107809.
   L_{\bar{x}}^2-431667. L_{\bar{x}}\right) 
   -34.2113 L_{\bar{x}}^2+\left(\bar{x}\right)^2
   \left(-5.38358 L_{\bar{x}}^5-180.755 L_{\bar{x}}^4-3771.6 L_{\bar{x}}^3-43136.2
   L_{\bar{x}}^2-262007. L_{\bar{x}}\right)-103.526 L_{\bar{x}} 
   +\bar{x} \left(1.85185
   L_{\bar{x}}^5+1.08025 L_{\bar{x}}^4-37.8328 L_{\bar{x}}^3-107.055 L_{\bar{x}}^2+85.0903
   L_{\bar{x}}+2491.42\right)+x \big(10.2642 L_x^5 
   +65.8827 L_x^4+175.497
   L_x^3-49.2838 L_x^2-1754.58 L_x-692466.\big)-1.98519 L_x^5+5.08025
   L_x^4-171.088 L_x^3 
   -103.238 L_x^2-2331.02 L_x+\frac{-177.716
   L_x-1109.28}{x}+1274.6 x^5+4328.89 x^4 \,,  
\end{dmath}

\begin{dmath}
I^{(3)}_{qg}(x) = N_f^2 \bigg\{61.0677\, +\bar{x}^3 \left(1.29041 L_{\bar{x}}^3-10.4998 L_{\bar{x}}^2+27.9136 L_{\bar{x}}\right)+\bar{x}^2 \left(-1.87711 L_{\bar{x}}^3-7.79618 L_{\bar{x}}^2-4.36051 L_{\bar{x}}\right)-0.987654 L_{\bar{x}}^3-4.34568 L_{\bar{x}}^2-8.2963 L_{\bar{x}}+\bar{x} \left(-0.987654 L_{\bar{x}}^3-5.53086 L_{\bar{x}}^2-12.2469 L_{\bar{x}}-13.2448\right)+0.0202177 x^6+0.107642 x^5+0.828892 x^4+21.886 x^3-26.2477 x^2-44.418 x-\frac{27.2797}{x}\bigg\}+N_f \bigg\{-617.49\, +x^3 \left(208.859 L_x^5+10.557 L_x^4+8704.26 L_x^3+1416.74 L_x^2+43827.5 L_x+34568.\right)+x^2 \left(-2.85131 L_x^5-105.472 L_x^4-1575.99 L_x^3-11432.7 L_x^2-36449.3 L_x+31014.7\right)+x \left(0.474074 L_x^5+1.11934 L_x^4+25.2949 L_x^3-8.30824 L_x^2+333.065 L_x-28583.3\right)-0.948148 L_x^5-2.83128 L_x^4-39.4623 L_x^3-104.954 L_x^2-206.825 L_x+\frac{66.855 L_x+299.169}{x}+\bar{x}^3 \left(197.15 L_{\bar{x}}^4-535.653 L_{\bar{x}}^3+8354.24 L_{\bar{x}}^2-17173.5 L_{\bar{x}}\right)+\bar{x}^2 \left(1.34535 L_{\bar{x}}^4-64.5016 L_{\bar{x}}^3-1685.89 L_{\bar{x}}^2-12486. L_{\bar{x}}\right)+2.88066 L_{\bar{x}}^4+28.3567 L_{\bar{x}}^3+127.182 L_{\bar{x}}^2+392.738 L_{\bar{x}}+\bar{x} \left(2.88066 L_{\bar{x}}^4+48.3073 L_{\bar{x}}^3+198.787 L_{\bar{x}}^2+471.463 L_{\bar{x}}+467.691\right)-147.604 x^6+1147.94 x^5-37319.1 x^4\bigg\}-5656.94\, +x^3 \left(304.814 L_x^5+710.583 L_x^4+16404.3 L_x^3+6246.83 L_x^2+151636. L_x+35826.5\right)+x^2 \left(-0.756674 L_x^5-25.0335 L_x^4-244.749 L_x^3+596.718 L_x^2+20659.8 L_x+30141.2\right)+x \left(-13.949 L_x^5+3.72016 L_x^4-169.304 L_x^3+1422.91 L_x^2+285.265 L_x+20018.4\right)+10.5877 L_x^5-58.7654 L_x^4+890.817 L_x^3-1109.35 L_x^2+10247.7 L_x+\frac{923.18 L_x^2+2860.53 L_x+9121.84}{x}+\bar{x}^3 \left(89.4404 L_{\bar{x}}^5+121.323 L_{\bar{x}}^4+3663.99 L_{\bar{x}}^3-409.556 L_{\bar{x}}^2+32059. L_{\bar{x}}\right)+\bar{x}^2 \left(-5.31508 L_{\bar{x}}^5-114.333 L_{\bar{x}}^4-990.403 L_{\bar{x}}^3-3650.08 L_{\bar{x}}^2-4439.25 L_{\bar{x}}\right)-2.46914 L_{\bar{x}}^5-30.5761 L_{\bar{x}}^4-210.131 L_{\bar{x}}^3-843.608 L_{\bar{x}}^2-2040.62 L_{\bar{x}}+\bar{x} \left(-2.46914 L_{\bar{x}}^5-58.3951 L_{\bar{x}}^4-435.152 L_{\bar{x}}^3-1600.97 L_{\bar{x}}^2-3150.68 L_{\bar{x}}-1728.79\right)-348.194 x^6+3445.64 x^5-94871.7 x^4 \,,
\end{dmath}
\begin{dmath}
I^{(3)}_{gg}(x)=-6265.65 x^6+66758.4 x^5-1.06264\times 10^6 x^4+\left(378.83 L_x^5+24477.3 L_x^4+63961.5 L_x^3+546586. L_x^2+910511. L_x+495749.\right) x^3+\left(57.2841 L_x^5+1945.83 L_x^4+27126.6 L_x^3+189429. L_x^2+615133. L_x+439670.\right) x^2+\left(-100.8 L_x^5-496. L_x^4-1570.17 L_x^3+7123.33 L_x^2+13857.3 L_x+19813.6\right) x+28.8 L_x^5-62. L_x^4+3055.28 L_x^3-176. L_{\bar{x}}^3+2488.89 L_x^2-839.824 L_{\bar{x}}^2+\bigg\{-0.0357895 x^6+0.0966543 x^5-4.13448 x^4+\left(0.13079 L_x^4-0.38099 L_x^3+4.56176 L_x^2-9.74592 L_x+21.6304\right) x^3+\left(-0.00273785 L_x^4-0.619734 L_x^3-2.15326 L_x^2-20.9999 L_x+148.982\right) x^2+\left(0.0987654 L_x^4-1.61317 L_x^3-25.3847 L_x^2-155.483 L_x-118.21\right) x+0.0987654 L_x^4-1.21811 L_x^3-19.2242 L_x^2+0.888889 L_{\bar{x}}^2-20.4598 \frac{1}{(\bar{x})}_+ -102.721 L_x+6.22222 L_{\bar{x}}+\bar{x}^2 \left(0.000979432 L_{\bar{x}}^3+1.21259 L_{\bar{x}}^2-8.48683 L_{\bar{x}}\right)+\bar{x} \left(0.197531 L_{\bar{x}}^3-3.2716 L_{\bar{x}}^2-17.8909 L_{\bar{x}}-121.663\right)+\bar{x}^3 \left(0.438091 L_{\bar{x}}^3+0.29313 L_{\bar{x}}^2+0.409493 L_{\bar{x}}\right)+41.2796\, -\frac{60.5749}{x}\bigg\} N_f^2+315.306 \left[\frac{1}{\bar{x}}\right]_++44694. L_x+1142.11 L_{\bar{x}}+\bar{x}^2 \left(-231.508 L_{\bar{x}}^3-499.573 L_{\bar{x}}^2+11535.3 L_{\bar{x}}\right)+\bar{x} \left(-200. L_{\bar{x}}^3-3.56745 L_{\bar{x}}^2-10.1778 L_{\bar{x}}-23676.7\right)+\bar{x}^3 \left(2652.95 L_{\bar{x}}^3-969.004 L_{\bar{x}}^2+51848.5 L_{\bar{x}}\right)+\bigg\{-65.3387 x^6+1878.62 x^5-64513.2 x^4+\left(426.686 L_x^5+47.588 L_x^4+19161.8 L_x^3-7284.34 L_x^2+136127. L_x+25910.7\right) x^3+\left(-4.21446 L_x^5-150.199 L_x^4-2113.28 L_x^3-13513. L_x^2-27711.8 L_x+22497.\right) x^2+\left(5.67407 L_x^5+50.2346 L_x^4+377.276 L_x^3+915.648 L_x^2+274.685 L_x+22082.4\right) x-3.79259 L_x^5-19.358 L_x^4-254.888 L_x^3+10.6667 L_{\bar{x}}^3-1313.87 L_x^2+64.5363 L_{\bar{x}}^2+347.079 \frac{1}{(\bar{x})}_+-3835.18 L_x-43.8845 L_{\bar{x}}+\bar{x}^3 \left(-83.5926 L_{\bar{x}}^4+181.9 L_{\bar{x}}^3-3263.87 L_{\bar{x}}^2+6458.66 L_{\bar{x}}\right)+\bar{x} \left(-0.493827 L_{\bar{x}}^4-3.45679 L_{\bar{x}}^3+14.1101 L_{\bar{x}}^2+470.287 L_{\bar{x}}+4782.37\right)+\bar{x}^2 \left(1.57238 L_{\bar{x}}^4+69.5168 L_{\bar{x}}^3+968.989 L_{\bar{x}}^2+5625.16 L_{\bar{x}}\right)-9860.15\, +\frac{176.957 L_x+788.476}{x}\bigg\} N_f+31575.9\, +\frac{2077.15 L_x^2+7128.59 L_x+23355.4}{x} \,.
\end{dmath}
\end{dgroup}
\subsubsection{Numerical fit for TMD FFs}

Following the same approach, we give in this subsection the results for TMD FFs.
The one-loop scale-independent coefficient functions are given by
\begin{subequations}
\begin{align}
C^{(1)}_{qq}(z) = & z^3 \left(6.88266 L_z-7.0639\right)+z^2 \left(10.467 L_z-0.842818\right)+z \left(5.33333 L_z+5.33121\right)\nonumber \\
&+5.33333 L_z+8. \bar{z}-0.128696 z^6+0.818297 z^5-3.44742 z^4-5.33333 \,,  \\ 
C^{(1)}_{gq}(z) = &\frac{10.6667 L_z}{z}-10.6667 L_z+z \left(5.33333 L_z-5.33333\right)-8. \bar{z}+8. \,, \\
C^{(1)}_{qg}(z) =& z \left(2.\, -4. L_z\right)+z^2 \left(4.
   L_z-2.\right)+2. L_z \,,  \\
C^{(1)}_{gg}(z) =& -12.\, +z^3 \left(15.486 L_z-15.8938\right)+z^2
   \left(-0.449268 L_z-1.89634\right)\nonumber  
   \\ 
   &  -24.
   L_z+\frac{24. L_z}{z}-0.289566 z^6+1.84117
   z^5-7.7567 z^4+12. \bar{z}  \nonumber 
   \\
   & +z
   \left(48. L_z+11.9952\right) \,.
\end{align}
\end{subequations}
The two-loop scale-independent coefficient functions are given by
\begin{dgroup}
\begin{dmath}
C^{(2)}_{qq}(z) =C^{(2)}_{q^\prime q}(z)+ N_f \bigg\{5.53086 \frac{1}{(\bar{x})}_+ +z^3 \left(1.70397 L_z^2-4.64799 L_z+10.359\right)+z^2 \left(0.881779 L_z^2-8.99257 L_z-0.401235\right)+z \left(0.444444 L_z^2-0.888889 L_z-3.55556\right)+0.444444 L_z^2-8. L_z-1.67901 \bar{z}-0.00933333 z^6+0.104686 z^5-1.16427 z^4-1.97531\bigg\}+14.9267 \frac{1}{(\bar{x})}_+ +\bar{z}^3 \left(3.93463 L_{\bar{z}}-0.760825 L_{\bar{z}}^2\right)+\bar{z}^2 \left(2.42655 L_{\bar{z}}^2-12.8657 L_{\bar{z}}\right)+7.11111 L_{\bar{z}}^2-22.2222 L_{\bar{z}}+\bar{z} \left(-3.55556 L_{\bar{z}}^2+47.1111 L_{\bar{z}}+139.708\right)+z^3 \left(-9.26406 L_z^3-52.8068 L_z^2+31.8935 L_z-175.006\right)+z^2 \left(-22.9319 L_z^3-1.84413 L_z^2-1.58947 L_z+42.8001\right)+z \left(-5.18519 L_z^3+6. L_z^2-122.654 L_z+146.514\right)-5.18519 L_z^3-14.4444 L_z^2+75.5681 L_z+0.720394 z^6-7.22814 z^5+85.3763 z^4-260.245 \,,
\end{dmath}

\begin{dmath}
C^{(2)}_{\bar{q}^\prime q}(z) =C^{(2)}_{q^\prime q}(z) \,,
\end{dmath}

\begin{dmath}
C^{(2)}_{\bar{q} q}(z) =C^{(2)}_{q^\prime q}(z)+z^2 \left(-1.2533 L_z^3+26.1088 L_z^2+241.727 L_z+645.84\right)+z \left(1.33333 L_z^3+0.888889 L_z^2-1.33333 L_z+18.5241\right)-1.33333 L_z^3-3.55556 L_z^2-8.44444 L_z-0.444444 \bar{z}-1.48376 z^6+20.64 z^5-282.996 z^4-10.8899+z^3 \left(47.853 L_z^3-13.7843 L_z^2+664.436 L_z-389.634\right) \,,
\end{dmath}

\begin{dmath}
C^{(2)}_{q^\prime q}(z) = -48.773\, +z^3 \left(0.0245052 L_z^3+0.125079 L_z^2-1.08241 L_z+3.45972\right)+z^2 \left(0.000177368 L_z^3-5.32882 L_z^2-5.58902 L_z+11.417\right)+z \left(4.88889 L_z^3-7.33333 L_z^2-18.6667 L_z+35.4396\right)+4.88889 L_z^3-7.33333 L_z^2-48. L_z+\frac{7.11111 L_z^2+3.55556 L_z-1.45999}{z}+1.33333 \bar{z}-0.000213858 z^6+0.00372439 z^5-0.0868925 z^4 \,,
\end{dmath}

\begin{dmath}
C^{(2)}_{g q}(z) = 1290.49\, +\bar{z}^3 \left(-3.84057 L_{\bar{z}}^3+37.6298 L_{\bar{z}}^2-86.9353 L_{\bar{z}}\right)+\bar{z}^2 \left(2.86982 L_{\bar{z}}^3+23.1076 L_{\bar{z}}^2-66.3682 L_{\bar{z}}\right)+1.48148 L_{\bar{z}}^3-4.44444 L_{\bar{z}}^2-48.3094 L_{\bar{z}}+\bar{z} \left(1.48148 L_{\bar{z}}^3+21.3333 L_{\bar{z}}^2-33.1982 L_{\bar{z}}-441.625\right)+z^3 \left(-5.92385 L_z^3-22.5299 L_z^2-110.802 L_z+15.1969\right)+z^2 \left(-0.212249 L_z^3-5.70494 L_z^2-67.3286 L_z+304.102\right)+z \left(-89.1852 L_z^3+84.8889 L_z^2+151.253 L_z-1484.97\right)-45.6296 L_z^3+21.3333 L_z^2+357.939 L_z+\frac{-106.667 L_z^3-282.667 L_z^2+156.728 L_z-152.722}{z}-0.587738 z^6-3.60247 z^5+53.0513 z^4 \,,
\end{dmath}

\begin{dmath}
C^{(2)}_{qg}(z) =-33.5701\, +N_f \bigg\{ \bar{z}^3 \left(2.042 L_{\bar{z}}-0.179298 L_{\bar{z}}^2\right)+\bar{z}^2 \left(0.670481 L_{\bar{z}}^2+0.602593 L_{\bar{z}}\right)+0.333333 L_{\bar{z}}^2+1.11111 L_{\bar{z}}+\bar{z} \left(-0.666667 L_{\bar{z}}^2-1.55556 L_{\bar{z}}+4.87603\right)+z^3 \left(-0.179321 L_z^2+2.04201 L_z+0.514002\right)+z^2 \left(0.670483 L_z^2-8.28627 L_z-2.88119\right)+z \left(-0.666667 L_z^2+7.33333 L_z+7.50775\right)+0.333333 L_z^2-3.33333 L_z-0.015025 z^6+0.0450704 z^5-0.294577 z^4-6.09183\bigg\}+\bar{z}^3 \left(3.18953 L_{\bar{z}}^3-3.71065 L_{\bar{z}}^2+13.5993 L_{\bar{z}}\right)+\bar{z}^2 \left(-1.01091 L_{\bar{z}}^3-5.20986 L_{\bar{z}}^2+41.8832 L_{\bar{z}}\right)-0.555556 L_{\bar{z}}^3-3.5 L_{\bar{z}}^2+1.78267 L_{\bar{z}}+\bar{z} \left(1.11111 L_{\bar{z}}^3+8.33333 L_{\bar{z}}^2-17.232 L_{\bar{z}}-79.1583\right)+z^3 \left(-6.58896 L_z^3-12.6016 L_z^2-148.67 L_z+226.183\right)+z^2 \left(-9.99409 L_z^3-45.4536 L_z^2-32.8022 L_z-189.651\right)+z \left(66.8889 L_z^3+2.66667 L_z^2-147.41 L_z-10.8933\right)+8.55556 L_z^3-24.3333 L_z^2-97.2949 L_z+\frac{16. L_z^2+8. L_z-3.28497}{z}-0.7087 z^6-0.502913 z^5+33.3265 z^4 \,,  
\end{dmath} 

\begin{dmath}
C^{(2)}_{gg}(z) =N_f \bigg\{12.4444 \frac{1}{(\bar{x})}_+ -127.778\, -0.000223065 \bar{z}^3 L_{\bar{z}}-0.0000208355 \bar{z}^2 L_{\bar{z}}+2. L_{\bar{z}}+z^3 \left(1.76728 L_z^3-4.55758 L_z^2+15.8982 L_z+0.643843\right)+z^2 \left(0.0299539 L_z^3+7.78684 L_z^2-25.7082 L_z+49.0399\right)+z \left(9.77778 L_z^3-19.3333 L_z^2-78. L_z+83.3332\right)+9.77778 L_z^3+7.33333 L_z^2-50. L_z+\frac{-0.888889 L_z^2-27.1111 L_z+3.4321}{z}+37.5556 \bar{z}-0.00782998 z^6+0.153818 z^5-3.92838 z^4\bigg\}+33.585\frac{1}{(\bar{x})}_+ +1795.49\, +\bar{z}^3 \left(0.0225989 L_{\bar{z}}^2+287.323 L_{\bar{z}}\right)+\bar{z}^2 \left(66.3821 L_{\bar{z}}^2+5.53377 L_{\bar{z}}\right)+36. L_{\bar{z}}^2-6. L_{\bar{z}}+\bar{z} \left(-18. L_{\bar{z}}^2+216. L_{\bar{z}}+444.823\right)+z^3 \left(-967.096 L_z^3-103.327 L_z^2-11404.1 L_z+4605.87\right)+z^2 \left(-21.4751 L_z^3-591.494 L_z^2-3594.61 L_z-9731.14\right)+z \left(-744. L_z^3-75. L_z^2+357.518 L_z-2387.27\right)-132. L_z^3+33. L_z^2+1292.74 L_z+\frac{-240. L_z^3-660. L_z^2+62.259 L_z-296.958}{z}+24.3517 z^6-407.965 z^5+5728.73 z^4  \,.
\end{dmath}
\end{dgroup}
The three-loop scale-independent coefficient functions are given by
\begin{dgroup}
\begin{dmath}
C^{(3)}_{qq}(z) =C^{(3)}_{q^\prime q}(z)+ N_f \bigg\{154.257 \frac{1}{(\bar{x})}_+  +\bar{z}^3 \left(1.03702 L_{\bar{z}}^3+0.00482051 L_{\bar{z}}^2-1.98623 L_{\bar{z}}\right)+\bar{z}^2 \left(-0.691297 L_{\bar{z}}^3-5.00561 L_{\bar{z}}^2-0.834261 L_{\bar{z}}\right)-2.107 L_{\bar{z}}^3-10.3704 L_{\bar{z}}^2+13.8356 L_{\bar{z}}+\bar{z} \left(1.0535 L_{\bar{z}}^3-3.55556 L_{\bar{z}}^2-26.922 L_{\bar{z}}+57.6679\right)+z^3 \left(-19.7036 L_z^4+33.1654 L_z^3-444.349 L_z^2+661.632 L_z-1870.9\right)+z^2 \left(-1.9698 L_z^4+30.6275 L_z^3+191.742 L_z^2+337.809 L_z+2071.18\right)+z \left(-0.148148 L_z^4-12.0933 L_z^3+50.6066 L_z^2+131.174 L_z-355.483\right)-0.148148 L_z^4+4.56516 L_z^3+115.265 L_z^2-458.905 L_z+0.5958 z^6-10.0011 z^5+300.139 z^4+138.622\bigg\}+N_f^2 \bigg\{-9.09324 \frac{1}{(\bar{x})}_++z^3 \left(1.13739 L_z^3-1.11292 L_z^2+8.49447 L_z+0.132844\right)+z^2 \left(0.911883 L_z^3-3.47057 L_z^2+5.15991 L_z-2.6942\right)+z \left(0.460905 L_z^3-0.855967 L_z^2-2.50206 L_z-0.131794\right)+0.460905 L_z^3-2.43621 L_z^2+8.82305 L_z-7.57543 \bar{z}-0.0139459 z^6+0.212281 z^5-3.95786 z^4+9.22493\bigg\}+140.136 \frac{1}{(\bar{x})}_+ +\bar{z}^3 \left(144.25 L_{\bar{z}}^4-566.24 L_{\bar{z}}^3+6271.93 L_{\bar{z}}^2-15644.6 L_{\bar{z}}\right)+\bar{z}^2 \left(-4.10249 L_{\bar{z}}^4-121.769 L_{\bar{z}}^3-1804.41 L_{\bar{z}}^2-12411.1 L_{\bar{z}}\right)+34.7654 L_{\bar{z}}^3+5.09037 L_{\bar{z}}^2-1826.42 L_{\bar{z}}+\bar{z} \left(-42.6667 L_{\bar{z}}^3+32.8626 L_{\bar{z}}^2+1865.35 L_{\bar{z}}-104.524\right)+z^3 \left(55.5684 L_z^5+409.681 L_z^4+1340.71 L_z^3+9286.53 L_z^2-212.8 L_z+19119.9\right)+z^2 \left(12.7948 L_z^5-7.34615 L_z^4-652.115 L_z^3-4213.45 L_z^2-9887.78 L_z+17171.7\right)+z \left(0.330864 L_z^5-14.4444 L_z^4+235.053 L_z^3-739.672 L_z^2+998.485 L_z-22860.2\right)+0.330864 L_z^5-1.73663 L_z^4-235.432 L_z^3-1059.64 L_z^2+3285.53 L_z-53.6084 z^6+420.432 z^5-16201.7 z^4+133.444 \,, 
\end{dmath}

\begin{dmath}
C^{(3)}_{\bar{q} q}(z) = C^{(3)}_{\bar{q}^\prime q}(z)+ N_f \bigg\{\bar{z}^3 \left(-10.7901 L_{\bar{z}}^3-1.1804 L_{\bar{z}}^2-172.111 L_{\bar{z}}\right)+\bar{z}^2 \left(-0.301929 L_{\bar{z}}^3-6.5035 L_{\bar{z}}^2-51.1538 L_{\bar{z}}\right)+\bar{z} \left(0.395062 L_{\bar{z}}+0.460905\right)+z^3 \left(-17.0335 L_z^4-137.582 L_z^3-499.752 L_z^2-1860.45 L_z-759.677\right)+z^2 \left(-1.91048 L_z^4-4.52615 L_z^3-95.3246 L_z^2-571.103 L_z-524.745\right)+z \left(0.773663 L_z^4-1.62414 L_z^3-7.72423 L_z^2+14.3251 L_z-277.011\right)-0.773663 L_z^4+1.16324 L_z^3+3.64193 L_z^2+13.9018 L_z+15.9632 z^6-137.732 z^5+1650.43 z^4+32.7672\bigg\}+\bar{z}^3 \left(1380.49 L_{\bar{z}}^4-3405.35 L_{\bar{z}}^3+54960. L_{\bar{z}}^2-107954. L_{\bar{z}}\right)+\bar{z}^2 \left(-26.476 L_{\bar{z}}^4-909.761 L_{\bar{z}}^3-12653. L_{\bar{z}}^2-84025.1 L_{\bar{z}}\right)+\bar{z} \left(-4.98979 L_{\bar{z}}-13.046\right)+z^3 \left(-486.111 L_z^5+2695.54 L_z^4-19026.4 L_z^3+99236. L_z^2-179433. L_z+137119.\right)+z^2 \left(15.2308 L_z^5+265.21 L_z^4+2452.17 L_z^3+12096.8 L_z^2-16346.8 L_z+124275.\right)+z \left(-2.35556 L_z^5-14.8148 L_z^4+109.379 L_z^3+142.666 L_z^2-100.216 L_z-192966.\right)+2.35556 L_z^5+24.4527 L_z^4-16.7701 L_z^3+103.935 L_z^2+333.13 L_z-1141.38 z^6+6288.46 z^5-73997. z^4+422.31 \,, 
\end{dmath}

\begin{dmath}
C^{(3)}_{q^\prime q}(z) = C^*_{q^\prime q}(z) + \frac{d^{ABC} d_{ABC}}{32 N_c} C_{d33}(z) \,,
\end{dmath}

\begin{dmath}
C^{(3)}_{\bar{q}^\prime q}(z) = C^*_{q^\prime q}(z) - \frac{d^{ABC} d_{ABC}}{32 N_c} C_{d33}(z) \,,
\end{dmath}

\begin{dmath}
C^*_{q^\prime q}(z)= N_f \bigg\{27.0915\, +\bar{z}^3 \left(0.193659 L_{\bar{z}}^3+0.293846 L_{\bar{z}}^2-2.24651 L_{\bar{z}}\right)+\bar{z}^2 \left(0.00288809 L_{\bar{z}}^3-1.70852 L_{\bar{z}}^2-7.99319 L_{\bar{z}}\right)+\bar{z} \left(-0.0987654 L_{\bar{z}}^3-0.493827 L_{\bar{z}}^2-1.28966 L_{\bar{z}}-3.86138\right)+z^3 \left(0.151889 L_z^4+0.157662 L_z^3-0.309824 L_z^2+6.09781 L_z+1.00545\right)+z^2 \left(-0.00133511 L_z^4-3.60493 L_z^3+16.2546 L_z^2-24.7872 L_z+12.0106\right)+z \left(0.790123 L_z^4-4.6749 L_z^3+23.4989 L_z^2+10.1958 L_z-30.7157\right)+0.790123 L_z^4-13.5638 L_z^3-2.87145 L_z^2+11.7244 L_z+\frac{6.32099 L_z^3+4.74074 L_z^2+5.47355 L_z-6.50358}{z}-0.0100925 z^6+0.0433735 z^5-2.92155 z^4\bigg\}+2356.76\, +\bar{z}^3 \left(1.99803 L_{\bar{z}}^4+0.201226 L_{\bar{z}}^3+60.9292 L_{\bar{z}}^2+14.0251 L_{\bar{z}}\right)+\bar{z}^2 \left(-0.026975 L_{\bar{z}}^4+1.5332 L_{\bar{z}}^3-1.66342 L_{\bar{z}}^2-74.0435 L_{\bar{z}}\right)+\bar{z} \left(0.246914 L_{\bar{z}}^4-0.938272 L_{\bar{z}}^3-4.14606 L_{\bar{z}}^2+61.3677 L_{\bar{z}}+90.0722\right)+z^3 \left(1.44997 L_z^5+5.483 L_z^4+123.168 L_z^3+176.819 L_z^2+923.553 L_z+28.2894\right)+z^2 \left(-0.00167084 L_z^5+22.4717 L_z^4+1.00054 L_z^3-146.27 L_z^2+790.245 L_z+5.17563\right)+z \left(-43.2296 L_z^5+85.358 L_z^4-173.688 L_z^3-919.477 L_z^2+3982.24 L_z-2387.44\right)+5.03704 L_z^5-14.7654 L_z^4+79.7975 L_z^3-990.195 L_z^2+273.338 L_z+\frac{-54.5185 L_z^4-353.975 L_z^3-346.753 L_z^2-131.107 L_z+712.406}{z}-3.27908 z^6+33.6575 z^5-745.562 z^4 \,,
\end{dmath}

\begin{dmath}
C_{d33}(z) = \bar{z}^2 \left(-0.307397 L_{\bar{z}}^2-38.4701 L_{\bar{z}}\right)+\bar{z} \left(-1.54596 L_{\bar{z}}^2+16.3038 L_{\bar{z}}+62.8465\right)+z^3 \left(2.97919 L_z^5+34.253 L_z^4+132.475 L_z^3+595.614 L_z^2+587.898 L_z+520.873\right)+z^2 \left(0.0487062 L_z^5-34.2258 L_z^4+134.207 L_z^3+383.35 L_z^2+246.025 L_z+472.994\right)+z \left(2.66667 L_z^4-260.948 L_z^3+967.077 L_z^2-1766.62 L_z+1419.94\right)+7.46667 L_z^5-6.22222 L_z^4-10.3848 L_z^3+706.144 L_z^2+521.357 L_z+4.63572 z^6-11.1449 z^5-872.326 z^4-1534.97+\bar{z}^3 \left(-0.269815 L_{\bar{z}}^2-28.5219 L_{\bar{z}}\right) \,,
\end{dmath}

\begin{dmath}
C^{(3)}_{g q}(z) = N_f \bigg\{-1108.48\, +\bar{z}^3 \left(7.58584 L_{\bar{z}}^4-31.4672 L_{\bar{z}}^3+228.754 L_{\bar{z}}^2-476.836 L_{\bar{z}}\right)+\bar{z}^2 \left(-0.940842 L_{\bar{z}}^4-14.0264 L_{\bar{z}}^3-53.8666 L_{\bar{z}}^2-174.207 L_{\bar{z}}\right)-0.411523 L_{\bar{z}}^4-1.86557 L_{\bar{z}}^3+31.6117 L_{\bar{z}}^2+126.856 L_{\bar{z}}+\bar{z} \left(-0.411523 L_{\bar{z}}^4-6.1454 L_{\bar{z}}^3-1.90267 L_{\bar{z}}^2+162.885 L_{\bar{z}}+913.267\right)+z^3 \left(4.57512 L_z^5+11.1861 L_z^4+203.498 L_z^3+290.533 L_z^2+1265.34 L_z+630.954\right)+z^2 \left(-0.0370741 L_z^5-1.40262 L_z^4-28.1512 L_z^3-174.684 L_z^2-95.9495 L_z+559.385\right)+z \left(-4.02963 L_z^5+6.02469 L_z^4+24.2085 L_z^3-459.232 L_z^2+915.724 L_z+599.07\right)+8.05926 L_z^5-24.4938 L_z^4-133.619 L_z^3-1101.17 L_z^2+1025.8 L_z+\frac{-36.0823 L_z^4-70.5844 L_z^3-30.0038 L_z^2-1644.91 L_z+531.011}{z}-3.68167 z^6+46.0164 z^5-1188.42 z^4\bigg\}-16659.8\, +\bar{z}^3 \left(35.4193 L_{\bar{z}}^5-219.487 L_{\bar{z}}^4+1485.12 L_{\bar{z}}^3-4170.57 L_{\bar{z}}^2+10384.9 L_{\bar{z}}\right)+\bar{z}^2 \left(-5.35238 L_{\bar{z}}^5-42.4256 L_{\bar{z}}^4+5.94854 L_{\bar{z}}^3-133.051 L_{\bar{z}}^2-1489.25 L_{\bar{z}}\right)-2.46914 L_{\bar{z}}^5+7.94239 L_{\bar{z}}^4+115.51 L_{\bar{z}}^3-515.669 L_{\bar{z}}^2-1954.74 L_{\bar{z}}+\bar{z} \left(-2.46914 L_{\bar{z}}^5-23.1687 L_{\bar{z}}^4+119.033 L_{\bar{z}}^3+1075.92 L_{\bar{z}}^2-2726.04 L_{\bar{z}}-17725.8\right)+z^3 \left(-76.1738 L_z^5-9.10623 L_z^4-2287.06 L_z^3-7866.64 L_z^2-5529.82 L_z-18309.8\right)+z^2 \left(0.537045 L_z^5+45.4989 L_z^4+819.376 L_z^3+722.523 L_z^2+15767.9 L_z-16612.4\right)+z \left(518.199 L_z^5-1121.6 L_z^4+829.631 L_z^3+21325.1 L_z^2-33619.9 L_z+29876.3\right)-89.9951 L_z^5+149.728 L_z^4+1805.41 L_z^3+20822. L_z^2-2827.94 L_z+\frac{384. L_z^5+3473.78 L_z^4+5160.83 L_z^3+6813.26 L_z^2+37045.7 L_z+7355.78}{z}+49.7303 z^6-749.182 z^5+14561.3 z^4 \,,
\end{dmath}

\begin{dmath}
C^{(3)}_{qg}(z) = -199.178 z^6+216.114 z^5-7090.88 z^4+\left(-156.025 L_z^5+166.04 L_z^4-4677.54 L_z^3+2353.55 L_z^2-15669. L_z+452.173\right) z^3+\left(10.1476 L_z^5+308.931 L_z^4+318.491 L_z^3+6648.7 L_z^2+38837.8 L_z+74.1398\right) z^2+\left(-395.365 L_z^5-435.299 L_z^4-651.612 L_z^3-873.892 L_z^2+8857.9 L_z-3975.65\right) z-0.925926 L_{\bar{z}}^5+16.8741 L_z^5-12.0833 L_{\bar{z}}^4-25.9012 L_z^4-54.8189 L_{\bar{z}}^3-255.16 L_z^3-28.4699 L_{\bar{z}}^2-4389.82 L_z^2+\bigg\{-0.000633714 z^6+0.0260241 z^5-0.665263 z^4+\left(-0.0765408 L_z^3+1.70812 L_z^2-8.03237 L_z+9.12771\right) z^3+\left(1.03454 L_z^3-5.86883 L_z^2+7.22527 L_z-4.90518\right) z^2+\left(-1.03704 L_z^3+5.25926 L_z^2-4.39289 L_z-9.7382\right) z-0.37037 L_{\bar{z}}^3+0.518519 L_z^3-1.85185 L_{\bar{z}}^2-1.85185 L_z^2+1.26572 L_{\bar{z}}+\bar{z}^2 \left(-0.73545 L_{\bar{z}}^3-1.69507 L_{\bar{z}}^2+8.70289 L_{\bar{z}}\right)+\bar{z}^3 \left(0.330843 L_{\bar{z}}^3-1.24575 L_{\bar{z}}^2+1.0939 L_{\bar{z}}\right)+\bar{z} \left(0.740741 L_{\bar{z}}^3+2.59259 L_{\bar{z}}^2-7.12403 L_{\bar{z}}-6.15555\right)-0.0998535 L_z+9.56988\bigg\} N_f^2+275.019 L_{\bar{z}}+\bar{z}^2 \left(-2.93658 L_{\bar{z}}^5-70.6885 L_{\bar{z}}^4-754.615 L_{\bar{z}}^3-4257.26 L_{\bar{z}}^2-13872.4 L_{\bar{z}}\right)+\bar{z} \left(1.85185 L_{\bar{z}}^5+28.7346 L_{\bar{z}}^4+139.144 L_{\bar{z}}^3-109.864 L_{\bar{z}}^2-1688.1 L_{\bar{z}}-2655.53\right)+\bar{z}^3 \left(88.7527 L_{\bar{z}}^5+138.689 L_{\bar{z}}^4+3798.44 L_{\bar{z}}^3+4087.64 L_{\bar{z}}^2+17552.1 L_{\bar{z}}\right)+1017.76 L_z+\bigg\{-21.5862 z^6+150.038 z^5-4826.58 z^4+\left(26.6804 L_z^5+4.22518 L_z^4+1004.59 L_z^3+571.353 L_z^2+4318.31 L_z+5231.6\right) z^3+\left(-0.395692 L_z^5-18.1614 L_z^4-227. L_z^3-1356.54 L_z^2-6051.6 L_z+4714.16\right) z^2+\left(3.02222 L_z^5-8.2716 L_z^4-63.1193 L_z^3-254.718 L_z^2+89.1164 L_z-4904.57\right) z-1.51111 L_z^5+1.08025 L_{\bar{z}}^4+0.765432 L_z^4+11.251 L_{\bar{z}}^3+0.806584 L_z^3+26.1788 L_{\bar{z}}^2+273.617 L_z^2-30.318 L_{\bar{z}}+\bar{z} \left(-2.16049 L_{\bar{z}}^4-24.428 L_{\bar{z}}^3-37.4317 L_{\bar{z}}^2+179.106 L_{\bar{z}}+330.174\right)+\bar{z}^2 \left(1.41617 L_{\bar{z}}^4-0.661483 L_{\bar{z}}^3-339.004 L_{\bar{z}}^2-2561.81 L_{\bar{z}}\right)+\bar{z}^3 \left(34.8293 L_{\bar{z}}^4-104.9 L_{\bar{z}}^3+1456.48 L_{\bar{z}}^2-2987.62 L_{\bar{z}}\right)+99.119 L_z-424.185\, +\frac{13.5638 L_z^3-9.74486 L_z^2+1.61518 L_z-10.5514}{z}\bigg\} N_f+9611.66\, +\frac{-122.667 L_z^4-814.222 L_z^3-1026.22 L_z^2-442.914 L_z+1652.6}{z} \,,
\end{dmath}

\begin{dmath}
C^{(3)}_{gg}(z) =13587.6 z^6-182029. z^5+4.31788\times 10^6 z^4+\left(-17646.9 L_z^5-32932.7 L_z^4-895202. L_z^3-506008. L_z^2-6.67683\times 10^6 L_z-2.23289\times 10^6\right) z^3+\left(111.781 L_z^5+4555.84 L_z^4+63933.2 L_z^3+367397. L_z^2+694258. L_z-1.99903\times 10^6\right) z^2+\left(4334.4 L_z^5+2099. L_z^4+1739.67 L_z^3+121402. L_z^2-10723.5 L_z+190560.\right) z-244.8 L_z^5+380. L_z^4+176. L_{\bar{z}}^3+3912.39 L_z^3+839.824 L_{\bar{z}}^2+61497.1 L_z^2+\bigg\{-0.0586466 z^6-0.0523985 z^5+3.93521 z^4+\left(-0.853738 L_z^4+3.18919 L_z^3-22.3459 L_z^2+51.4518 L_z-104.039\right) z^3+\left(0.00929978 L_z^4-3.60852 L_z^3-3.74294 L_z^2+104.196 L_z+35.241\right) z^2+\left(1.18519 L_z^4-25.5144 L_z^3+51.8078 L_z^2+58.689 L_z-146.248\right) z+1.18519 L_z^4-27.0947 L_z^3-0.888889 L_{\bar{z}}^2+5.70903 L_z^2-20.4598 \frac{1}{(\bar{x})}_+ -6.22222 L_{\bar{z}}+\bar{z}^2 \left(0.00178891 L_{\bar{z}}^3-1.13567 L_{\bar{z}}^2-2.99102 L_{\bar{z}}\right)+\bar{z} \left(0.197531 L_{\bar{z}}^3-0.604938 L_{\bar{z}}^2-16.9247 L_{\bar{z}}-99.3166\right)+\bar{z}^3 \left(0.618545 L_{\bar{z}}^3+1.03642 L_{\bar{z}}^2-0.613943 L_{\bar{z}}\right)+113.908 L_z+240.465\, +\frac{0.263374 L_z^3+1.5144 L_z^2+37.1139 L_z-7.39814}{z}\bigg\} N_f^2+315.306 \frac{1}{(\bar{x})}_+ -7159.3 L_{\bar{z}}+\bar{z}^3 \left(-4055.59 L_{\bar{z}}^3-2593.37 L_{\bar{z}}^2-48610.9 L_{\bar{z}}\right)+\bar{z} \left(-376. L_{\bar{z}}^3-1179.39 L_{\bar{z}}^2+12927.7 L_{\bar{z}}+3854.88\right)+\bar{z}^2 \left(234.799 L_{\bar{z}}^3-891.557 L_{\bar{z}}^2-27609.5 L_{\bar{z}}\right)-656.548 L_z+\bigg\{-366.473 z^6+5156.93 z^5-148894. z^4+\left(771.541 L_z^5+751.33 L_z^4+35545.6 L_z^3+5384.26 L_z^2+246456. L_z+81634.5\right) z^3+\left(-6.90626 L_z^5-256.962 L_z^4-3539.19 L_z^3-23870. L_z^2-52024.3 L_z+71845.1\right) z^2+\left(-114.03 L_z^5+344.525 L_z^4+647.313 L_z^3-3967.84 L_z^2+7537.44 L_z-8883.28\right) z+32.237 L_z^5-60.2469 L_z^4-10.6667 L_{\bar{z}}^3-287.664 L_z^3-64.5363 L_{\bar{z}}^2-6169.48 L_z^2+347.079 \frac{1}{(\bar{x})}_+ +61.0452 L_{\bar{z}}+\bar{z}^2 \left(-1.53589 L_{\bar{z}}^4-72.1586 L_{\bar{z}}^3-964.26 L_{\bar{z}}^2-5462.05 L_{\bar{z}}\right)+\bar{z} \left(-0.493827 L_{\bar{z}}^4+4.24691 L_{\bar{z}}^3+66.1118 L_{\bar{z}}^2+0.917326 L_{\bar{z}}+2403.55\right)+\bar{z}^3 \left(96.2988 L_{\bar{z}}^4-145.492 L_{\bar{z}}^3+3741.37 L_{\bar{z}}^2-6474.06 L_{\bar{z}}\right)-5069.6 L_z+100.426\, +\frac{-74.0741 L_z^4+136.494 L_z^3+909.237 L_z^2-3714.97 L_z+905.703}{z}\bigg\} N_f-145882.\, +\frac{864. L_z^5+8008. L_z^4+15340.7 L_z^3+23082.4 L_z^2+76348.7 L_z+19111.9}{z}\,.
\end{dmath}
\end{dgroup}

\section{Small $x$ expansion of unpolarized TMD coefficients and  resummation for TMD FFs}
\label{sec:small-x-expansion}

In this section we give the results for TMD PDFs and FFs expanded in high energy limit, namely $x,z \to 0$. A striking difference between space-like TMD PDFs and time-like TMD FFs is that there is only single logarithmic enhancement in each order of perturbative expansion in TMD PDFs, while in TMD FFs it becomes double logarithmic enhancement. We also resum the small-$z$ logarithms in TMD FFs to Next-to-Next-to-Leading Logarithmic~(NNLL) accuracy in this section.

\subsection{Small-$x$ expansion of unpolarized TMD PDFs}
\label{sec:small-x-expansion-1}
Using the analytic expression we obtained, it is straightforward to obtain the small-$x$ expansion. At leading power in the expansion, the results read
\begin{align}
x I^{(2)}_{qq} (x)=&x  I^{(2)}_{qq'} (x)= x I^{(2)}_{q\bar{q}} (x)=x I^{(2)}_{q\bar{q}'} (x)= 2 C_F T_F\left(  \frac{172}{27}-\frac{8 \zeta _2}{3} \right) \,,  
\nn
\\
x I^{(3)}_{qq} (x) =&x I^{(3)}_{qq'}(x) =x I^{(3)}_{q\bar{q}} (x)= x I^{(3)}_{q\bar{q}'} (x)\nonumber
\\
 =& 2 T_F\bigg[  \left(\frac{208 \zeta _2}{9}+\frac{32 \zeta _3}{3}-\frac{17152}{243}\right) C_A C_F \ln x 
 + \left(-16 \zeta _2 +\frac{512}{9} \zeta _3 +\frac{32}{3} \zeta _4 -\frac{269}{9} \right) C_F^2
 \nn\\
   +&\left(\frac{12008 \zeta _2}{81}+120 \zeta _3+\frac{920 \zeta
   _4}{9}-\frac{456266}{729}\right) C_A C_F
   +\left(-\frac{32 \zeta _2}{9}-\frac{64 \zeta
   _3}{9}+\frac{16928}{729}\right) C_F N_f T_F
  \bigg] \label{eq:smallx2}\,. 
\end{align}

\begin{align}
x I^{(2)}_{qg}(x) =& 2 C_A T_F \left(  \frac{172}{27}-\frac{8 \zeta _2}{3} \right)  \,, \nonumber  
\\  
x I^{(3)}_{qg}(x) =& 2 T_F \bigg[
   \left(\frac{208 \zeta _2}{9}+\frac{32 \zeta
   _3}{3}-\frac{17152}{243}\right) C_A^2 \ln x 
 + \left(\frac{160 \zeta _2}{27}-\frac{32 \zeta _3}{9}-\frac{3164}{729}\right) C_A N_f
   T_F  \nonumber 
   \\
   +&\left(-16 \zeta _2+\frac{512 \zeta _3}{9}+\frac{32 \zeta
   _4}{3}-\frac{269}{9}\right) C_A C_F   +\left(\frac{12536 \zeta
   _2}{81}+\frac{1096 \zeta _3}{9}+\frac{920 \zeta _4}{9}-\frac{470494}{729}\right) C_A^2
   \nn\\
   +&\left(-\frac{512 \zeta _2}{27}-\frac{64 \zeta _3}{9}+\frac{40184}{729}\right) C_F
   N_f T_F \bigg] \label{eq:smallx1} \,,   
\end{align}

\begin{align}
x I^{(2)}_{gq} (x)=&
C_A C_F \bigg[\frac{88 }{3}\zeta_2+48 \zeta_3
-\frac{3160}{27}\bigg]+\frac{448 }{27}C_F N_f T_F\,,
   \nn\\
x I^{(3)}_{gq}(x) =&
C_A^2 C_F \bigg[64 \zeta_3 \ln ^2x+\bigg(-\frac{7504}{27} \zeta_2-80 \zeta_3
-\frac{392 }{3}\zeta_4+\frac{75584}{81}\bigg) \ln x
   \nn\\
   -&\frac{103304 }{81}\zeta_2
   +\frac{320}{3} \zeta_2 \zeta_3 -\frac{1504}{3} \zeta_3-\frac{12436 }{9}\zeta_4
   -\frac{4208 }{3}\zeta_5+\frac{333613}{54}\bigg]\nn\\
   +& C_A C_F^2
   \bigg[-\frac{512}{3} \zeta_3 \zeta_2
   +88 \zeta_2-672 \zeta_3+368 \zeta_4+\frac{2432 }{3}\zeta_5-\frac{1105}{6}\bigg]
   \nn\\
   +&C_A C_F N_f T_F
   \bigg[\bigg(-\frac{512}{27} \zeta_2
   +64 \zeta_3+\frac{1424}{81}\bigg) \ln
   x-\frac{2432}{27} \zeta_2+\frac{832 }{9}\zeta_3+16 \zeta_4
      \nn\\
   +&\frac{68548}{243}\bigg]
   +C_F^3 \bigg[192 \zeta_3 \zeta_2-104 \zeta_2
  +592 \zeta_3-392 \zeta_4-640 \zeta_5+\frac{467}{3}\bigg]
   \nn\\
   +&C_F^2 N_f
   T_F \bigg[\bigg(\frac{1024 }{27}\zeta_2-\frac{128}{3} \zeta_3
  -\frac{19040}{243}\bigg) \ln x+\frac{9776 }{81}\zeta_2-\frac{1696 }{9}\zeta_3
   -32 \zeta_4
    \nn\\
   -&\frac{139334}{729}\bigg]
   +C_F N_f^2 T_F^2
   \bigg[-\frac{128}{3} \zeta_3-\frac{7424}{243}\bigg]\,.
\label{eq:smallxg1}
\end{align}

\begin{align}
x I^{(2)}_{gg} (x)=&
C_A^2 \bigg[\frac{88 }{3}\zeta_2+48 \zeta_3-\frac{3160}{27}\bigg]+\frac{484
  }{27} C_A N_f T_F-\frac{8 }{3}C_F N_f T_F\,,
   \nn\\
x I^{(3)}_{gg} (x)=&
C_A^3 \bigg[64 \zeta_3 \ln ^2x+\bigg(-\frac{7504}{27} \zeta_2-\frac{176 }{3}\zeta_3
   -\frac{392 }{3}\zeta_4+\frac{75584}{81}\bigg) \ln x
      \nn\\
   -&\frac{112928}{81} \zeta_2
   +128 \zeta_2 \zeta_3-\frac{1792}{3} \zeta_3-1452 \zeta_4-1232 \zeta_5
   +\frac{1572769}{243}\bigg]
   \nn\\
   +&C_A^2 N_f T_F \bigg[\bigg(-\frac{512
}{27}\zeta_2
+\frac{320 }{3}\zeta_3+\frac{1568}{81}\bigg) \ln x-\frac{16288 }{81}\zeta_2
    -\frac{512 }{9}\zeta_3
   \nn\\
   +&\frac{1040}{9} \zeta_4
    +\frac{535048}{729}\bigg]
   + C_A C_F N_f T_F
   \bigg[\bigg(\frac{1024 }{27}\zeta_2-128 \zeta_3-\frac{19904}{243}\bigg) \ln
   x
   \nn\\
   +&\frac{10144 }{27}\zeta_2+\frac{256 \zeta_3}{9}-\frac{1376}{9} \zeta_4
   -\frac{836194}{729}\bigg]+C_A N_f^2 T_F^2 \bigg[-\frac{128 }{9}\zeta_3-\frac{40160}{729}\bigg]
   \nn\\
   +&C_F^2 N_f T_F \bigg[\frac{160
   }{9}\zeta_2-\frac{896}{9} \zeta_3
   +\frac{1024}{9} \zeta_4
   +12\bigg]+C_F
   N_f^2 T_F^2 \bigg[\frac{35776}{729}-\frac{512}{9} \zeta_3\bigg]\,.
\label{eq:smallxg2}
\end{align}
We note that a LL prediction for the small-$x$ expansion has been given in \cite{Marzani:2015oyb}. After fixing a typo in that paper, we find full agreement with its LL prediction for both quark and gluon TMD PDFs at small $x$.~\footnote{We thank Simone Marzani for communicating with us the typo in Eq.~(40) of \cite{Marzani:2015oyb}.} It would be very interesting to extend the formalism of \cite{Marzani:2015oyb} beyond LL and compare with the data presented here.

\subsection{Small-$z$ expansion of unpolarized TMD FFs}
\label{sec:small-x-expansion-2}

To facilitate small-$z$ resummation for TMD FFs, we shall consider the coefficient functions in flavor singlet sector below, since non-singlet TMD FFs are at most logarithmic divergent, but not power divergent in the $z\to 0$ limit. 
The flavor singlet~(denoted by a superscript$\ ^s$) coefficient functions can be written as a matrix,
\begin{align}
\widehat C^s(z) = 
  \begin{pmatrix}
 \widetilde{C}_{qq}(z) & 2 N_f C_{qg}(z)
\\
    C_{gq} (z) & C_{gg}(z)
  \end{pmatrix} \,,
\end{align}
where
\begin{equation}
\label{eq:singlet}
 \widetilde{C}_{qq}(z)=C_{qq}(z)+C_{\bar q q}(z)+ (N_f-1) (C_{q'q}(z)+ C_{\bar q' q}(z)) \,,
\end{equation}
and $C_{ij}(z)$ are scaleless coefficient functions as appeared in the solutions of RG equation \eqref{eq:RGST}.

In contrast to TMD PDFs, which contribute a single logarithm at each perturbative order in the  small-$x$,
TMD FFs in the singlet sector develop small-$z$ double logarithms,
\begin{align}
\label{eq:double-log}
\lim_{z\to 0} z \widehat C^s_{k j}(z)= \lim_{z \to 0} z \sum_{n=1}^\infty a_s^n \widehat C^{s(n)}_{k j}(z)   { \sim}   \sum_{n=1}^\infty a_s^n \bigg(  \sum_{m=1}^{2n} {\ln^{2n-m} z} \bigg)\,,
\end{align}
where $a_s=\alpha_s/(4 \pi)$ is our perturbative expansion parameter.
The small-$z$ data for quark fragmentation in the singlet sector are (non-singlet sector results are suppressed in the $z \to 0$ limit)
\begin{align}
z \widehat C^{s(1)}_{qq}(z)=&0\,,
\nn\\
z  \widehat C^{s(2)}_{qq}(z)=&2N_fC_F T_F 
\bigg(
\frac{32}{3}\ln^2z+\frac{16}{3}\ln z+\frac{16}{3}\zeta_2-\frac{296}{27}
\bigg)\,,
\nn\\
z \widehat C^{s(3)}_{qq}(z)=&
2N_fC_A C_F T_F
\bigg[
-\frac{736}{27}\ln^4 z-\frac{14336}{81}\ln^3 z+\bigg(-\frac{64}{9}\zeta_2-\frac{14632}{81}\bigg)\ln^2 z
+\bigg(\frac{256}{3}\zeta_3
\nn\\
-&\frac{3616}{27}\zeta_2+\frac{11312}{243}\bigg)\ln z
-\frac{1472}{9}\zeta_4+112\zeta_3-\frac{17600}{81}\zeta_2+\frac{512156}{729}
\bigg]
\nn\\
+&2N_fC_F^2 T_F
\bigg[
\frac{128}{3}\ln^2z+\bigg(-128\zeta_3+\frac{128}{9}\zeta_2+\frac{1288}{9}\bigg)\ln z
-\frac{608}{3}\zeta_4+\frac{32}{9}\zeta_3+\frac{6272}{27}\zeta_2
\nn\\
-&\frac{1262}{27}
\bigg]
+2 C_F N_f^2 T_F^2
\bigg[
\frac{512}{27}\ln^3z+\frac{128}{9}\ln^2z+
\bigg(-\frac{256}{9}\zeta_2+\frac{5120}{81}\bigg)\ln z
-\frac{64}{9}\zeta_2-\frac{5696}{729}
\bigg]\,.
\end{align}
\begin{align}
z \widehat C^{s(1)}_{gq}(z)=&8 C_F \ln z\,,
\nn\\
z \widehat C^{s(2)}_{gq}(z)=&
 C_A C_F\bigg[
 -\frac{80}{3}\ln^3 z-\frac{212}{3}\ln^2 z +(32 \zeta_2+12)\ln z -88\zeta_3-\frac{88}{3}\zeta_2+\frac{3128}{27}
  \bigg]
 \nn\\
 +&
 C_F^2\bigg[
 (48-64\zeta_2)\ln z
 \bigg]\,,
 \nn\\
 z \widehat C^{s(3)}_{gq}(z)=&
 C_A^2 C_F
 \bigg[
 32 \ln^5 z
 +\frac{7816}{27}\ln^4 z
 +\left(\frac{7376}{9}-\frac{2560}{9}\zeta_2\right)\ln^3 z
 +\bigg(-\frac{1984}{3}\zeta_2+\frac{1184}{3}\zeta_3
 \nn\\
 +&\frac{8608}{9}\bigg)\ln^2 z
 +\bigg(\frac{3776}{3}\zeta_2+\frac{3136}{3}\zeta_3+1408\zeta_4-\frac{123892}{81}\bigg)\ln z
 +\frac{7456 \zeta_5}{3}+\frac{608}{3}\zeta_2 \zeta_3
 \nn\\
 +&\frac{5870}{3}\zeta_4-\frac{7588}{9}\zeta_3+\frac{21944}{27}\zeta_2-\frac{3650707}{729}
 \bigg]
 +C_A C_F^2
 \bigg[
 \bigg(\frac{1792}{9}\zeta_2-\frac{1360}{9}\bigg)\ln^3 z
 \nn\\
 +&\bigg(-\frac{64}{3}\zeta_3+\frac{1888}{3}\zeta_2-\frac{1532}{3}\bigg)\ln^2 z
 +\bigg(\frac{488}{3}\zeta_4-\frac{80}{3}\zeta_3-\frac{1360}{3}\zeta_2+\frac{701}{9}\bigg)\ln z
 \nn\\
 +&\frac{6800}{3}\zeta_5+992 \zeta_3\zeta_2+\frac{830}{3}\zeta_4-1746\zeta_3-\frac{48568}{27}\zeta_2+\frac{10141}{9}
 \bigg]
 \nn\\
 +&C_A C_F N_f T_F
 \bigg[
-\frac{32}{27}\ln^4 z+ \frac{3712}{81}\ln^3 z+\bigg(\frac{896}{9}\zeta_2+\frac{152}{81}\bigg)\ln^2 z
+\bigg(-\frac{320}{3}\zeta_3
\nn\\
-&\frac{4384}{27}\zeta_2-\frac{67384}{243}\bigg)\ln z
+\frac{688}{9}\zeta_4+\frac{704}{3}\zeta_3+\frac{10864}{81}\zeta_2-\frac{50600}{729}
\bigg]
\nn\\
+&C_F^3 \bigg[
\bigg(\frac{128}{3}\zeta_3-64\zeta_2+\frac{208}{3}\bigg)\ln^2 z
+\bigg(\frac{1600}{3}\zeta_4-336\zeta_3-\frac{224}{3}\zeta_2-\frac{173}{3}\bigg)\ln z
\nn\\
-&416\zeta_5-\frac{2240}{3}\zeta_3\zeta_2+796\zeta_4+\frac{2888}{3}\zeta_3+608\zeta_2-\frac{4715}{3}
 \bigg]
 \nn\\
+&C_F^2 N_f T_F
\bigg[
-\frac{1024}{27}\ln^4z-\frac{4928}{27}\ln^3z+\bigg(-\frac{256}{3}\zeta_2-\frac{7184}{27}\bigg)\ln^2z
\nn\\
+&\bigg(-\frac{448}{3}\zeta_3-\frac{1280}{9}\zeta_2+\frac{2060}{27}\bigg)\ln z
-\frac{224}{3}\zeta_4-\frac{1664}{9}\zeta_3-\frac{5984}{27}\zeta_2+\frac{75770}{729}
\bigg]\,.
\end{align}
\begin{align}
z \widehat C^{s(1)}_{qg}(z)=&0\,,
\nn\\
z \widehat C^{s(2)}_{qg}(z)=&
2 N_f C_A T_F \bigg[\frac{32 }{3}\ln ^2z+\frac{16 }{3}\ln z+\frac{16}{3} \zeta_2
   -\frac{296}{27}\bigg]\,,
\nn\\
z \widehat C^{s(3)}_{qg}(z)=&
2 N_f C_A^2 T_F \bigg[\bigg(-\frac{832}{9} \zeta_2
-\frac{1160}{9}\bigg) \ln ^2z
+\bigg(-\frac{4640}{27} \zeta_2
+\frac{256}{3} \zeta_3+\frac{14360}{243}\bigg) \ln z
   \nn\\
   -&\frac{736}{27} \ln ^4z-\frac{14656}{81} \ln ^3z
   -\frac{2368}{27} \zeta_2+\frac{976}{9} \zeta_3
   -\frac{2432 }{9}\zeta_4+\frac{152392}{243}\bigg]
   \nn\\
   +&2 N_f C_A C_F T_F
   \bigg[\bigg(\frac{256 }{3}\zeta_2-\frac{64}{3}\bigg) \ln ^2z+\bigg(\frac{512 }{9}\zeta_2-128 \zeta_3+\frac{1000}{9}\bigg) \ln z
   \nn\\
   +&\frac{3040 }{27}\zeta_2+\frac{32
   }{9}\zeta_3-96 \zeta_4+\frac{514}{27}\bigg]
   +2 C_A N_f^2 T_F^2
   \bigg[\bigg(\frac{1136}{81}-\frac{512 }{27}\zeta_2\bigg) \ln z
   \nn\\
   +&\frac{896}{81} \ln
   ^3z
   -\frac{3328}{81} \ln ^2z-\frac{512 }{81}\zeta_2-\frac{64}{9} \zeta_3
   +\frac{256}{27}\bigg]
   +2 C_F N_f^2 T_F^2
   \bigg[\bigg(\frac{224}{3}
   \nn\\
   -&\frac{512 }{27}\zeta_2\bigg) \ln z
   +\frac{1280}{81} \ln
   ^3z+\frac{5120}{81} \ln ^2z
   -\frac{2048 }{81}\zeta_2+\frac{128}{9} \zeta_3
   +\frac{10304}{729}\bigg]\,.
\end{align}
\begin{align}
z \widehat C^{s(1)}_{gg}(z)=&8 C_A\ln z\,,
\nn\\
z \widehat C^{s(2)}_{gg}(z)=
&C_A^2 \bigg[\bigg(\frac{536}{9}-32 \zeta_2\bigg) \ln z-\frac{80}{3} \ln
   ^3z-\frac{220 }{3}\ln ^2z-\frac{88 }{3}\zeta_2-88 \zeta_3
   +\frac{3268}{27}\bigg]
   \nn\\
   +&C_A N_f T_F \bigg[-\frac{16}{3} \ln
   ^2z-\frac{184 }{9}\ln z+\frac{556}{27}\bigg]
   +C_F N_f T_F
   \bigg[\frac{32 }{3}\ln ^2z+\frac{16}{3} \ln z-\frac{1112}{27}\bigg]\,,
\nn\\
z \widehat C^{s(3)}_{gg}(z)=&C_A^3 \bigg[\bigg(\frac{57392}{81}-\frac{256}{3} \zeta_2\bigg) \ln
   ^3z
   +\bigg(-\frac{352 }{3}\zeta_2+416 \zeta_3+\frac{14792}{27}\bigg) \ln
   ^2z
      \nn\\
   +&\bigg(\frac{19984}{27} \zeta_2+\frac{5632}{9} \zeta_3+2104 \zeta_4
   -\frac{344864}{243}\bigg) \ln z+32 \ln ^5z+\frac{8008}{27} \ln
   ^4z
   \nn\\
   -&\frac{34640}{81} \zeta_2+448 \zeta_2 \zeta_3-\frac{14576 }{9}\zeta_3
   +3256 \zeta_4+4336 \zeta_5-\frac{1348136}{243}\bigg]
   \nn\\
   +&C_A^2 N_f
   T_F \bigg[\bigg(\frac{256 }{3}\zeta_2+\frac{22816}{81}\bigg) \ln
   ^2z+\bigg(-\frac{3616 }{27}\zeta_2+\frac{256 }{9}\zeta_3-\frac{46912}{81}\bigg)
   \ln z
   \nn\\
   +&\frac{352}{27} \ln ^4z+\frac{5152}{27} \ln ^3z-\frac{3136 }{81}\zeta_2
   +\frac{3152 }{9}\zeta_3+240 \zeta_4-\frac{230840}{729}\bigg]
   \nn\\
   +&
   C_A C_F N_f T_F \bigg[\bigg(-\frac{256}{9} \zeta_2
   -\frac{37072}{81}\bigg) \ln ^2z
   +\bigg(-\frac{3008}{27} \zeta_2
   -\frac{2560}{9}\zeta_3+\frac{92560}{243}\bigg) \ln z
   \nn\\
   -&\frac{1792}{27} \ln ^4z-\frac{29248
  }{81} \ln ^3z+\frac{3856 }{27}\zeta_2-\frac{1552 }{3}\zeta_3-\frac{8720}{9} \zeta_4+\frac{849464}{729}\bigg]
   \nn\\
   +&C_A N_f^2 T_F^2 \bigg[\frac{256 }{81}\ln
   ^3z+\frac{1472}{81} \ln ^2z+\frac{3776}{243} \ln z
   -\frac{128}{9} \zeta_3 -\frac{52256}{729}\bigg]
   \nn\\
   +&C_F^2 N_f T_F
   \bigg[\bigg(\frac{1024 }{9}\zeta_3-\frac{352}{3}\bigg) \ln z+\frac{64 }{3}\ln
   ^2z+\frac{2752 }{9}\zeta_3+\frac{832 }{3}\zeta_4-\frac{1684}{3}\bigg]
   \nn\\
   +&C_F
   N_f^2 T_F^2 \bigg[-\frac{512}{81} \ln ^3z-\frac{2944}{81} \ln
   ^2z+\frac{18560}{243} \ln z+\frac{256 }{9}\zeta_3+\frac{104512}{729}\bigg]\,.
\end{align}
We note that while both $\widehat C_{qi}^s$ and $\widehat C_{gi}^s$ has double logarithmic expansion in the small-$z$ limit, the power of leading logarithmic terms of $\widehat C_{qi}^s$ is lower by $1$ than the corresponding leading logarithmic terms of $\widehat C_{gi}^s$. 

\subsection{Resummation of small-$x$ logarithms for unpolarized TMD FFs }
\label{sec:resummation-small-x}

In this subsection, we shall derive the all-order resummation at NNLL accuracy~(resummation of the highest three logarithms) in Eq.~(\ref{eq:double-log}), following an idea proposed in \cite{Vogt:2011jv}. To this end, we begin with the unrenormalized version of collinear factorization formula in Eq.~\eqref{eq:mass-fac-form} for singlet TMD FFs~(see \eqref{eq:singlet} for the definition of singlet combination), 
\begin{align}
\label{eq:massfac}
 {\cal F}_{i/j}^s(z,\e) =\frac{1}{Z_j^B}  \frac{{{\cal F }}_{i/j}^{{s,\rm bare}}(z,\e)}{\mathcal{S}_{0 \rm b} }= \sum_k  d_{ik}^s  \otimes  {C_{kj}^s(z,\e)} \,,
\end{align}
where the convolution is in $z$.
Note that $\mathcal{F}_{i/j}^s(z,\epsilon)$ is a quantity to which the usual strong coupling renormalization, zero-bin subtraction and operator renormalization have been performed. However renormalization of collinear FFs has not been performed, which is why we still keep the $\e$ dependence in \eqref{eq:massfac}. From now on  we concentrate on the scale-independent part of the coefficient functions by setting all the scale logarithms to zero. We can do this because the scale logarithms depends either on anomalous dimension, which has no $z$ dependence, or on time-like splitting functions, whose small-$z$ behavior is known to NNLL accuracy~\cite{Kom:2012hd}.
It proves convenient to work in Mellin-$N$ space also, which is defined as
\begin{align}
  \label{eq:Mellindef}
  {\cal F}(\overline{N}, \e) = M[{\cal F}(z,\e)] :=  \int_0^1 dz \, z^{N-1} {\cal F}(z, \e)  \,,
\end{align}
where $\overline{N} = N - 1$. Small-$z$ logarithms becomes poles in $\bar N$ under Mellin transformation,
\begin{align}
M\left[\frac{1}{z}\ln^k z\right]\equiv\int_{0}^{1}dz\, z^{N-1}\frac{1}{z}\ln^k z=\frac{(-1)^k k!}{(N-1)^{k+1}}=\frac{(-1)^k k!}{\overline N^{k+1}}\,.
\end{align}
In Mellin space the unrenormalized collinear factorization formula in Eq.~(\ref{eq:massfac}) becomes
\begin{align}
\label{eq:massfacN}
\left(
\begin{array}{c}
 \mathcal{F}_{qi}^s(\overline{N},\e) \\
 \mathcal{F}_{gi}^s(\overline{N},\e) \\
\end{array}
\right) = 
 {\widehat d}^s(\overline{N},\e)\cdot \left(
\begin{array}{c}
 \widehat C_{qi}^s(\overline{N},\e) \\
 \widehat C_{gi}^s(\overline{N},\e) \\
\end{array}
\right) \,,
\end{align}
where
\begin{align}
  \label{eq:dmatrix}
  \widehat d^s(\overline{N},\e) = \left(
\begin{array}{cc}
\widehat d^s_{qq}(\overline{N},\e) & \widehat d^s_{qg}(\overline{N},\e) \\
 \widehat d^s_{gq}(\overline{N},\e) & \widehat d^s_{gg}(\overline{N},\e) \\
\end{array}\right)\,.
\end{align}

The collinear FFs in $\overline{\rm MS}$ scheme evolve with time-like splitting functions. In Mellin moment space it reads
\begin{align}
\label{eq:dglap}
\frac{d }{d \ln \mu^2} \widehat d^s(\overline{N},\e)   =
2 \widehat d^s(\overline{N},\e) \cdot \widehat \gamma^T(\overline{N})\,,
\end{align}
where $\widehat \gamma^T(\overline{N})$ is the time-like singlet splitting function in Mellin space. Its complete NNLO results can be found in \cite{Chen:2020uvt}, see also \cite{Mitov:2006ic,Moch:2007tx,Almasy:2011eq}.

The crucial observation of \cite{Vogt:2011jv} is that unrenormalized collinear functions have specific singular behavior in the limit of small-$z$ in dimensional regularization. In the case of TMD FFs, we can write down an general ansatz at small $z$,
\begin{align}
{\cal F}^{s(n)}_{g/i}(z,\e) =& \frac{1}{\epsilon^{2n-1}} \sum_{l=0}^{{n-1}} z^{-1-2(n-l) \epsilon} \big( \underbrace{c^{(1,l,n)}_{gi}}_{\text{LL}}+ \underbrace{\epsilon c^{(2,l,n)}_{gi}}_{\text{NLL}} +\underbrace{ \epsilon^2 c^{(3,l,n)}_{gi}}_{\text{NNLL}} + \dots\big) \,,
\nn\\
{\cal F}^{s(n)}_{q/i}(z,\e) =& \frac{1}{\epsilon^{2n-2}} \sum_{l=0}^{{n-2}} z^{-1-2(n-l) \epsilon} \big( \underbrace{c^{(1,l,n)}_{qi}}_{\text{LL}}+ \underbrace{\epsilon c^{(2,l,n)}_{qi}}_{\text{NLL}} +\underbrace{ \epsilon^2 c^{(3,l,n)}_{qi}}_{\text{NNLL}} + \dots\big) \,,
\label{eq:Fspace}
\end{align}
where $c_{gi}^{(1,l,n)}$ is the leading term  in the $\e$ expansion and small-$z$ expansion, whose knowledge correspond to LL resummation as labeled in \eqref{eq:Fspace}, and similarly for other terms. Precisely, for $\widehat{C}_{gi}^s(z)$ the LL series correspond to $\alpha_s^n \ln^{2 n -1}z$ terms, while NLL correspond to $\alpha_s^n \ln^{2n-2}z$, and NNLL to $\alpha_s^n \ln^{2n-3}z$. For $\widehat{C}_{qi}^s(z)$ the corresponding power of $\ln z$ is lower by $1$.
We have verified this general ansatz through explicit N$^3$LO calculation from its operator definition in \eqref{eq:FF_hadron_Frame} and \eqref{eq:partontohadron}. 
In Mellin space the corresponding ansatz reads
\begin{align}
\label{eq:ansantzNspace}
{\cal F}^{s(n)}_{g/i}(\overline{N},\e) =& \frac{1}{\epsilon^{2n-1}}  \sum_{l=0}^{{n-1}}   \frac{1}{\overline{N} -2 (n-l) \epsilon} \big( {c^{(1,l,n)}_{gi}}+ {\epsilon c^{(2,l,n)}_{gi}} +{ \epsilon^2 c^{(3,l,n)}_{gi}} +\dots\big)\,,
\nn\\
{\cal F}^{s(n)}_{q/i}(\overline{N},\e) =& \frac{1}{\epsilon^{2n-2}}  \sum_{l=0}^{{n-2}}  \frac{1}{\overline{N} -2 (n-l) \epsilon} \big( {c^{(1,l,n)}_{qi}}+ {\epsilon c^{(2,l,n)}_{qi}} +{ \epsilon^2 c^{(3,l,n)}_{qi}} +\dots\big)\,.
\end{align}
Equations ~\eqref{eq:Fspace} or \eqref{eq:ansantzNspace} provides a way to resum all the large logarithms of $z$. Specifically, if one knows all the $c_{gi}^{1,l,n}$ for all $l$ and $n$, then one can do LL resummation for ${\cal F}_{g/i}^{s}$, and similarly for NLL and NNLL resummation. In this paper instead of working out the constants for all $n$, we provide results for $n$ up to $15$, which is sufficient for phenomenological purpose. 

We note that the neglected terms in \eqref{eq:ansantzNspace} are higher orders in $\e$ and in $\overline{N}$. To facilitate easy extraction of the  constants, it is convenient to define a ``small-$z$'' weight:
\begin{align}
  \label{eq:weight}
  [\overline{N}] = 1 \,, \qquad [\e] = 1 \,, \qquad [\text{numbers}]=0 \,,
\end{align}
such that $c_{gi}^{(m,l,n)}$ corresponds to the weight $-(2n-1)-1+(m-1) = m -2 n -1$ terms in the small $\overline{N}$ expansion for ${\cal F}_{g/i}^{s(n)}$. For NNLL resummation, only  $m=1,2,3$ are needed. 

In order to determine the constants relevant for NNLL resummation in \eqref{eq:Fspace} for all $l$ and $n$, it turns out that only finite number of input is needed. Let us analyse Eq.~\eqref{eq:ansantzNspace} in more detail, taking $\mathcal{F}^{s(n)}_{g/i}(\overline{N},\epsilon)$ as an example. To LL accuracy, one need to determine $n$ unkown coefficients $c^{(1,l,n)}_{gi}$ with $l=0,\cdots n-1$. The lowest order of $\epsilon$ appearing in the ansatz $\mathcal{F}^{s(n)}_{g/i}(\overline{N},\epsilon)$ is $\epsilon^{-2n+1}$. Therefore expanding $\epsilon$ to order 
\begin{align}
(-2 n + 1)  + (n - 1) = -n   
\end{align}
gives us $n$ conditions and is enough to determine the $n$ unknowns at LL accuracy. To acheive NNLL accuracy, one only needs two more power of $\e$ expansion, that is we need to know $\mathcal{F}^{s(n)}_{g/i}(\overline{N},\epsilon)$ to order
$\epsilon^{-n+2}$. 
Similar analysis shows that we also need to know  $\mathcal{F}^{s(n)}_{q/i}(\overline{N},\epsilon)$ to order $\epsilon^{-n+2}$ to achive NNLL accuracy.

Having this important information in hand, let us contrentrate on the right hand side of Eq.~\eqref{eq:massfacN}, which generates the necessary $\e$ poles. The dimensionally regularized partonic FFs $\widehat d^s(\overline{N},\e)$ in $\overline{\text{MS}}$ scheme is purely divergent in $\epsilon$ and can be determined easily by solving Eq.~\eqref{eq:dglap} order by order in $\alpha_s$, 
\begin{align}
d \ln \widehat d^s (\overline{N},\e) = {d {a_s}} \frac{- \widehat \gamma^T(\overline{N})}{ {a_s} ( \e  + \sum_{n=0}^\infty {a_s}^{n+1} \beta_n )}\,.
\end{align}
where we have traded $d\ln \mu^2$ for $d a_s$ with the help of the $(4 -  2 \e)$-dimension beta function, 
\begin{align}
\frac{d {a_s}}{ d \ln \mu^2} = -2 \epsilon {a_s} -2 {a_s} \sum_{n=0}^\infty {a_s}^{n+1} \beta_n \,.
\end{align}
Assuming  $\widehat d^s(\overline{N},\e) = 1+ \sum_{k=1}^\infty a_s^{k}\, \widehat d_{k}^s$ and $\widehat \gamma^T(\overline{N}) = \sum_{k=0}^\infty {a_s}^{k+1} \widehat \gamma_{k}^{T}$, the solutions can be worked out order by order. For example, at first four orders we have
\begin{align}
\label{eq:ffCtofourthorder}
\widehat d_1^s =& -\frac{\widehat \gamma^T _0}{\epsilon }  \,,\quad \widehat d_2^s = \frac{-\widehat  d_1^s \cdot \widehat \gamma ^T_0-\widehat \gamma ^T_1}{2 \epsilon }+\frac{\beta _0 \widehat \gamma ^T_0}{2 \epsilon ^2} \,,
  \nn\\
\widehat  d_3^s =& \frac{\beta _0 \widehat d_1^s \cdot \widehat \gamma ^T_0+\beta _0 \widehat \gamma ^T_1+\beta _1 \widehat \gamma ^T_0}{3 \epsilon
   ^2}+\frac{-\widehat  d_1^s \cdot \widehat \gamma ^T_1- \widehat d_2^s\cdot\widehat \gamma ^T_0-\widehat \gamma ^T_2}{3 \epsilon }-\frac{\beta _0^2 \widehat \gamma
   ^T_0}{3 \epsilon ^3}  \,, \nonumber      \\ 
\widehat d_4^s =& -\frac{\beta _0 \left(\beta _0 \widehat d_1^s.\widehat \gamma^T_0+\beta _0 \widehat \gamma^T_1+2 \beta _1 \widehat \gamma
   ^T_0\right)}{4 \epsilon ^3} \nonumber  \\
   & +\frac{\beta_0 \widehat d_1^s.\widehat \gamma ^T_1+\beta _0 \widehat d_2^s.\widehat \gamma^T_0+\beta _1
   \widehat d_1^s.\widehat \gamma^T_0+\beta _0 \widehat \gamma ^T_2+\beta _2 \widehat \gamma ^T_0+\beta _1 \widehat \gamma ^T_1}{4 \epsilon
   ^2}  \nonumber \\
   & +\frac{-\widehat d_1^s.\widehat \gamma ^T_2-\widehat d_2^s.\widehat \gamma^T_1-\widehat d_3^s.\widehat \gamma^T_0-\widehat \gamma^T_3}{4 \epsilon
   }+\frac{\beta _0^3 \widehat \gamma ^T_0}{4 \epsilon ^4}\,. 
\end{align}
As explained above, to NNLL accuracy, we need to determine the coefficients of the pole terms in ${\cal F}_{g(q)/i}^{s(n)}$ to order $\e^{-n+2}$. Since the matching coefficients $\widehat C_{g(q)i}^s(\overline{N},\e)$ in \eqref{eq:massfacN} must be finite in the limit of $\e \to 0$, we need to determine the coefficients of the pole terms in $\widehat d_n^s$ also to order $\e^{-n+2}$. Also recall that $\widehat d_n^s$ are function of $\overline{N}$ and we are only interested in the small $\overline{N}$ limit. For NNLL resummation we only need to keep the terms in $\widehat d_n^s$ with ``small-$z$'' weight up to $2-2n$, see Eq.~\eqref{eq:weight}. We also note that the lowest weight term in $\widehat \gamma^T_n$ is $1-2 n$, corresponding to the leading $1/\overline{N}^{2n-1}$ pole. Using these information it can be shown that the required inputs for $\widehat d_n^s$ to achieve NNLL accuracy are $\beta_{0}$ and $\widehat \gamma^T_{0,1,2}$. One can check that this is indeed the case frm Eq.~\eqref{eq:ffCtofourthorder}, where explicit examples for $n \leq 4$ are shown.

Having known the general form of counter terms $d(\overline{N},\epsilon)$ to any order in $\alpha_s$, and the fact that the coefficient function $\widehat C_{g(q)i}^s$ is finite in the $\e \to 0$ limit, we can solve Eq.~\eqref{eq:massfacN} recursively to get the coefficient for the required pole terms in ${\cal F}_{g(q)/i}^s$, order by order in $\alpha_s$. In summary the input data we need to achive NNLL accuracy are 
\begin{align}
&\beta_0\,,  \nonumber \\ 
& \gamma^T_0\, \,, \gamma^T_1 \, \,, \gamma^T_2\,, \nonumber\\
&{\cal F}_{g(q)/i}^{s(0)} = 1 \, \,,  \mathcal{F}_{g(q)/i}^{s(1)} \text{ to } \epsilon^1 \, \,, \mathcal{F}_{g(q)/i}^{s(2)} \text{ to }   \epsilon^0  \,.
\end{align}
Note that although we have the explicit results for ${\cal F}_{g(q)/i}^{s(3)}$ from direct calculation, they are not needed for predicting the small-$z$ logarithms at NNLL accuracy. Rather they can be used as a check for our resummation. 


Following the approach outlined above, we have determined the coefficients of $\e$ poles in ${\cal F}_{g(q)/i}^{s(n)}$ up to $\e^{-n+2}$ for $n \leq 15$. From these coefficients we solve for the LL to NNLL constants defined in \eqref{eq:Fspace}. Expanding the results in $\e$ gives us the resummed NNLL series truncated at order $\alpha_s^{15}$, which should be sufficient for phenomenology. We have checked that a truncation of perturbative series at $\alpha_s^{15}$ leads to a less that $1\%$ relative uncertainty. The analytic expressions for the truncated resummed perturbative series can be found in the ancillary files of this paper.

At LL, we are able to find the generating function for the series
\begin{align}
\widehat {C}^{s}_{gq}(\overline{N})|_{\text{LL}} = \sum_{n=1}^\infty a_s^{n} \overline{N}^{-2 n} C_F C_A^{n-1} A_n\,,
\end{align}
with
\[
A_n = \frac{(-1)^n 2^{5 n} \Gamma \left(n+\frac{1}{4}\right)}{\Gamma \left(\frac{1}{4}\right) \Gamma (n+1)} \,.
\]
The series can be resummed analytically, leading to an closed form expression for LL results, 
\begin{align}
\widehat{C}^{s}_{gg} (\overline{N})|_{\text{LL}}=\frac{C_A}{C_F} \widehat {C}^{s}_{gq} (\overline{N})|_{\text{LL}} =\left(1+\frac{32 C_A a_s}{{\overline N}^2}\right)^{-1/4}-1\,.
\label{eq:LLres}
\end{align}
It's interesting to note that Eq.~\eqref{eq:LLres} coincides with that of the transverse coefficient functions for semi-inclusive $e^+e^-$ annihilation~\cite{Vogt:2011jv,Albino:2011si}.

\begin{figure}[ht!]
\centering
   \includegraphics[width=0.45\textwidth]{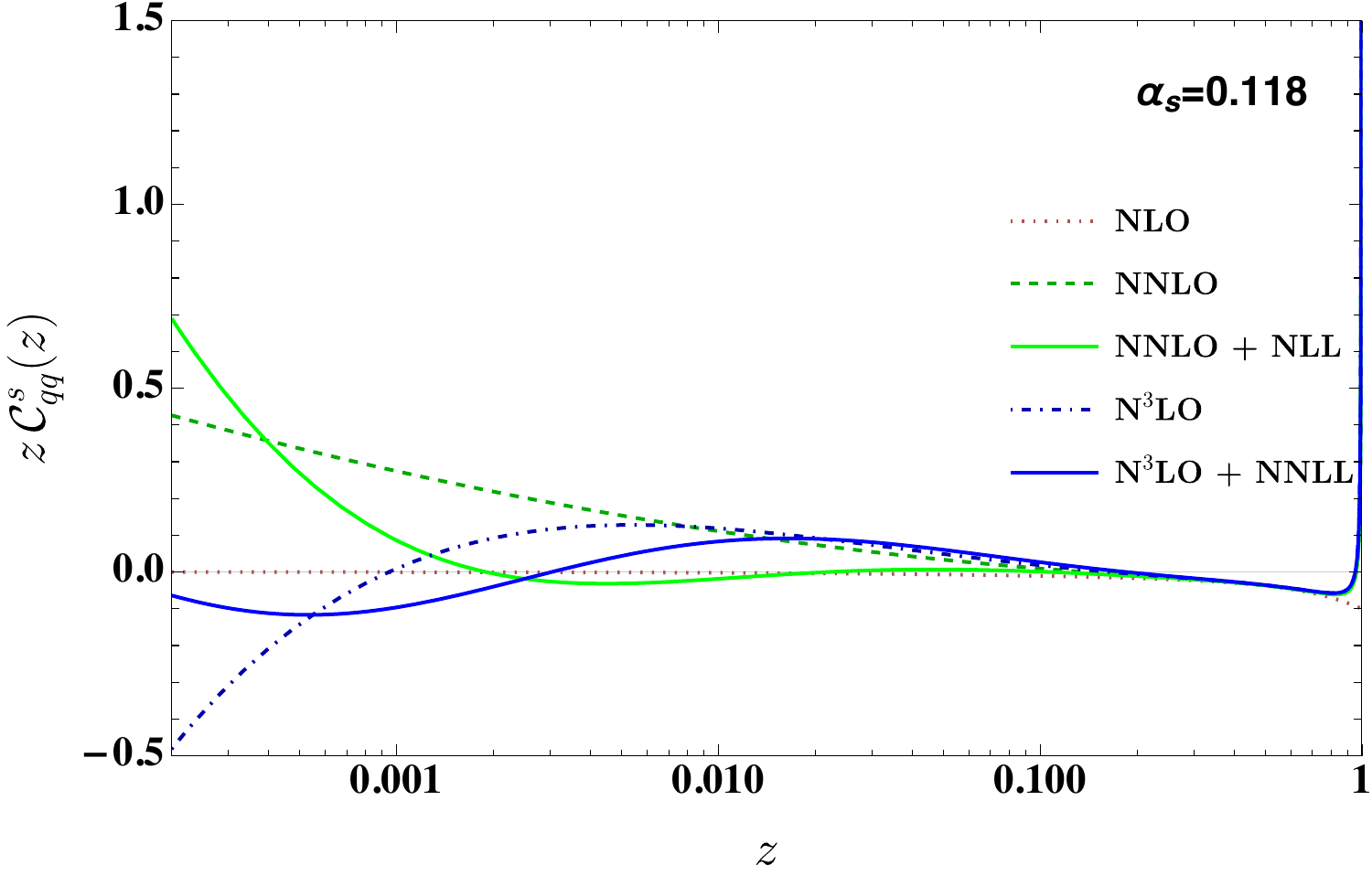}
    \includegraphics[width=0.45\textwidth]{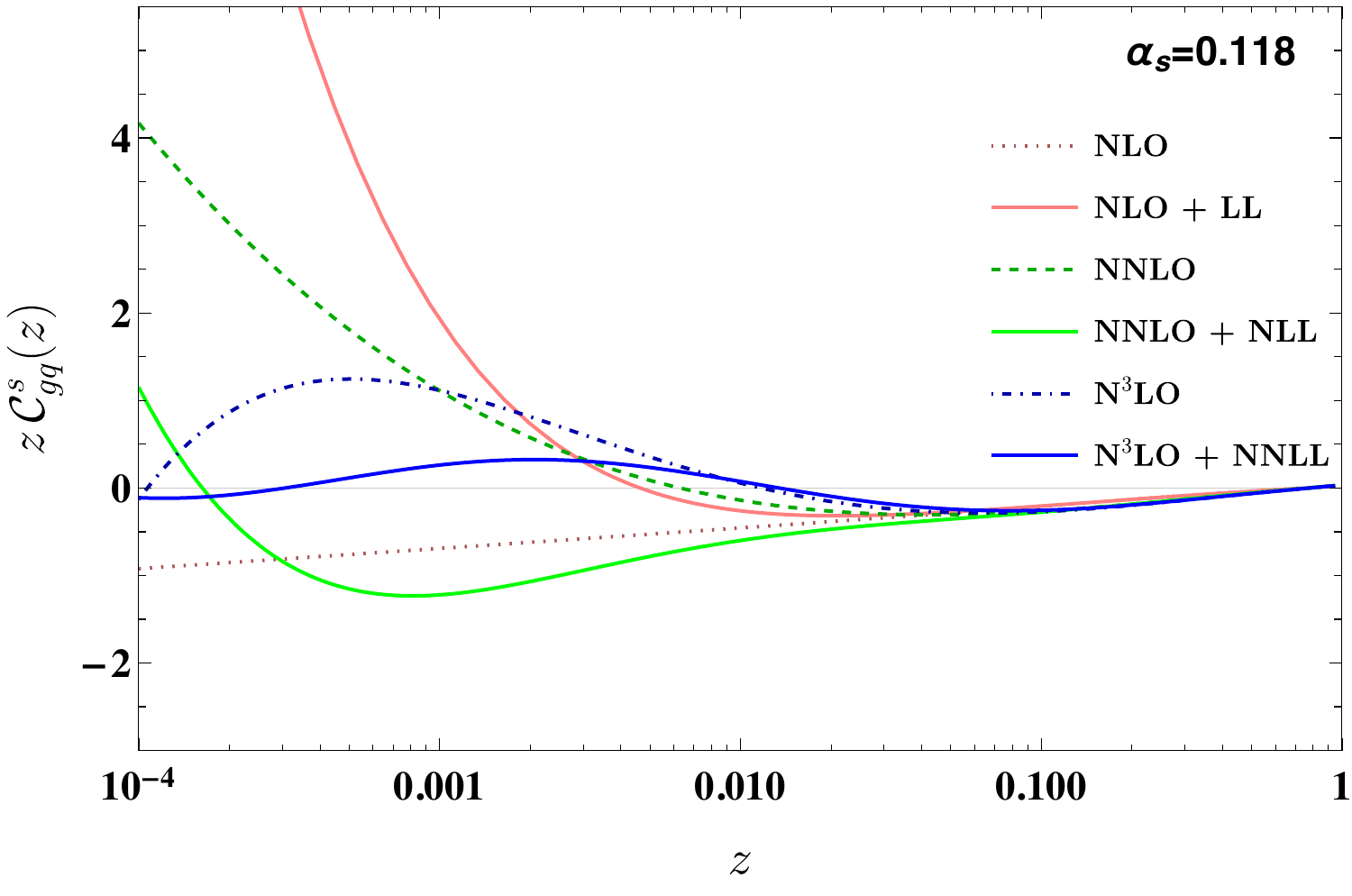}
    \caption{Coefficient functions for quark TMD FFs. Shown in the plots are fixed-order results at NLO, NNLO and  N$^3$LO, as well as adding the higher-order resummation contributions truncated to order $\alpha_s^{15}$.}
\label{fig:Cq2}
 \end{figure}
\begin{figure}[ht!]
\centering
   \includegraphics[width=0.45\textwidth]{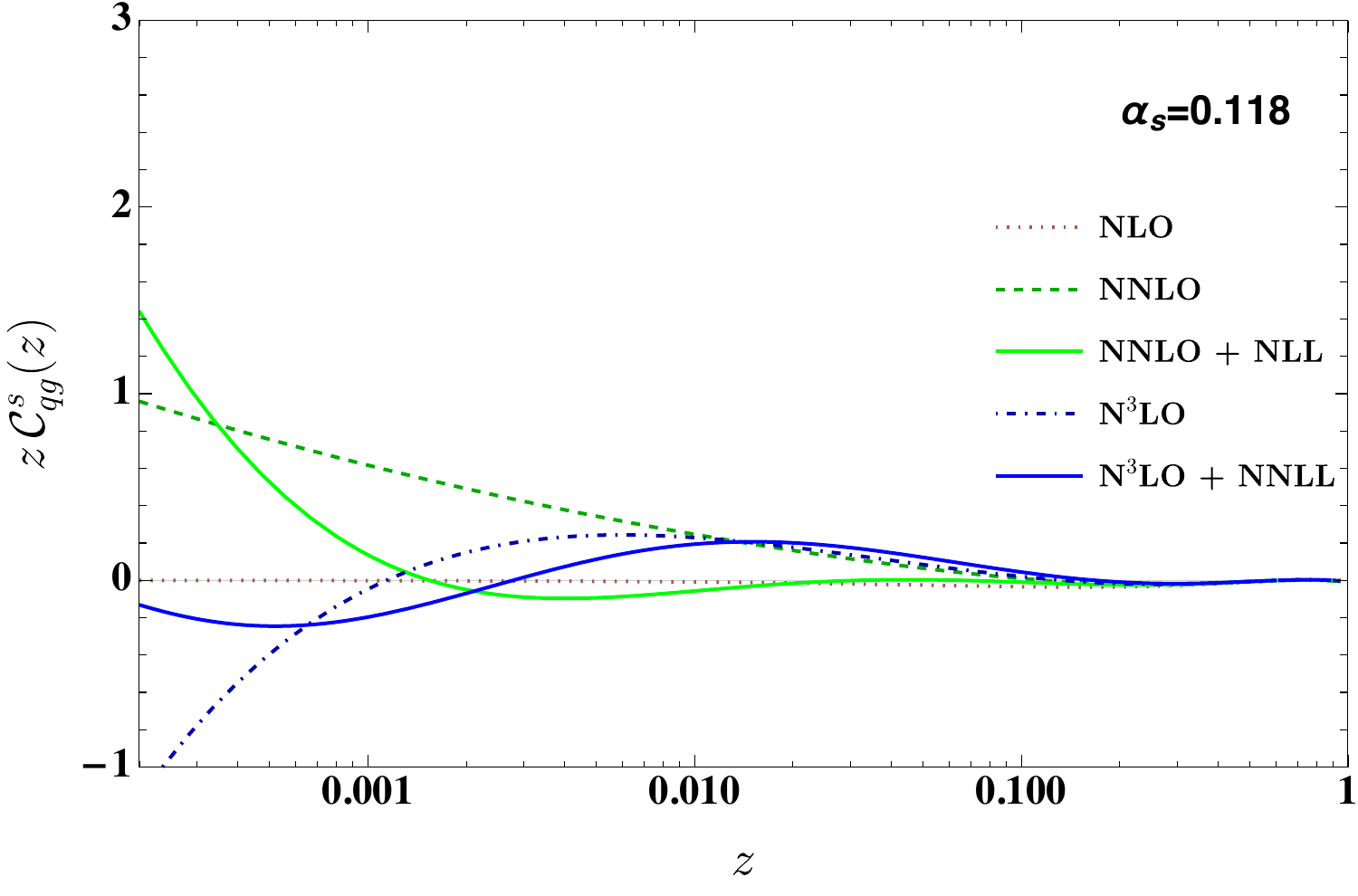}
    \includegraphics[width=0.45\textwidth]{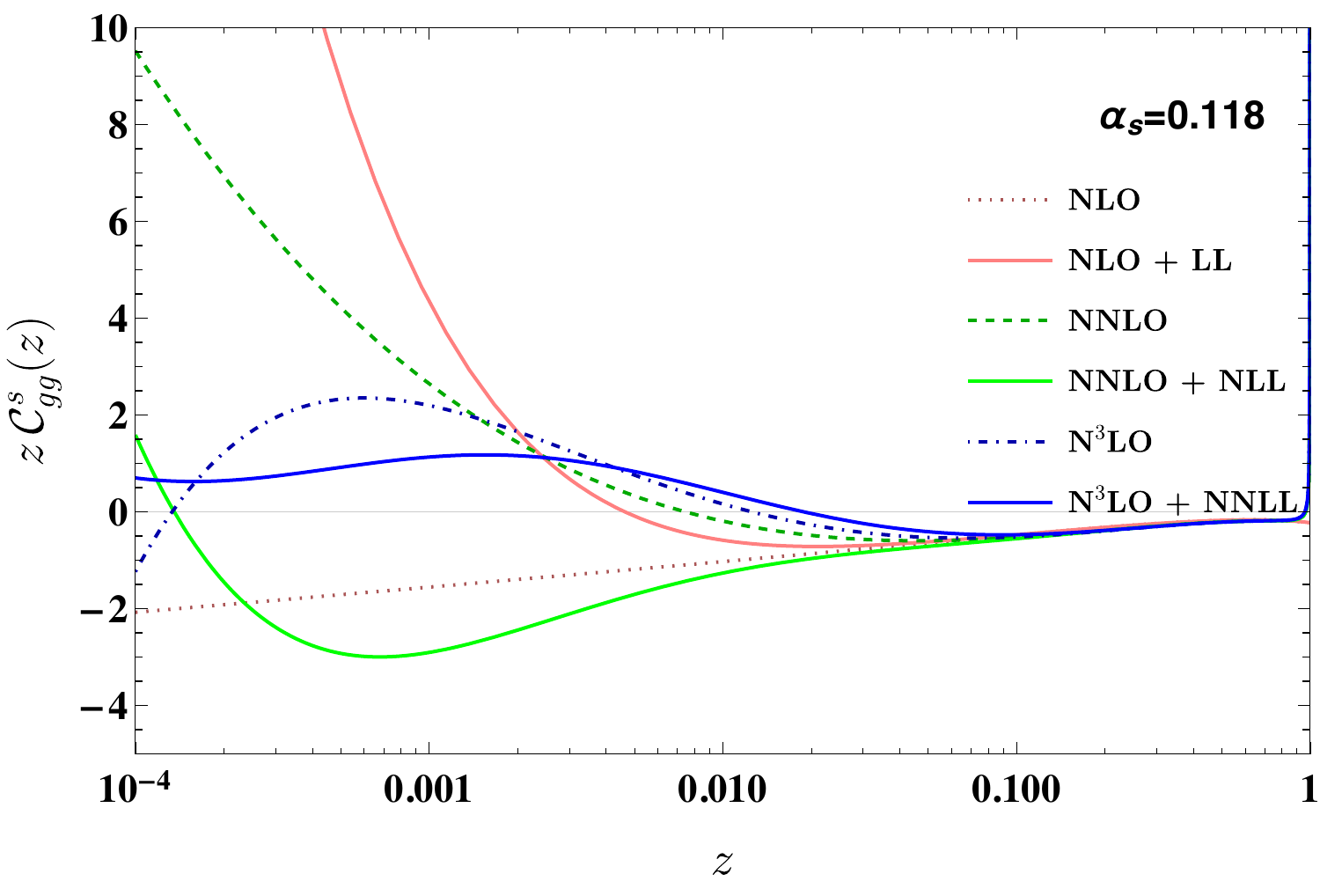}
    \caption{Coefficient functions for gluon TMD FFs. Shown in the plots are fixed-order results at NLO, NNLO and  N$^3$LO, as well as adding the higher-order resummation contributions truncated to order $\alpha_s^{15}$.}
\label{fig:Cg2}
 \end{figure}

In Fig.~\ref{fig:Cq2} and \ref{fig:Cg2} we plot the fixed-order coefficient functions and with different orders of small-$z$ resummation. We use $N_f = 5$ throughout the calculations. We note that even at N$^3$LO, the effects of resummation is important for $z < 10^{-2}$.


\section{Conclusion}
\label{sec:conclusion}

In summary we presented calculations for the unpolarized quark and gluon TMD PDFs and FFs at N$^3$LO in QCD. The unpolarized quark TMD PDFs at N$^3$LO have already been reported in Ref.~\cite{Luo:2019szz}. The rest of the calculations are new. Unpolarized quark and gluon TMD PDFs have also been calculated in \cite{Ebert:2020yqt} recently using an independent method, whose results are in full agreement with ours~\footnote{Except for a minor error in an anomalous color factor in \cite{Luo:2019szz}, which has been corrected in the arXiv version of that paper.}.

Our calculations for TMD PDFs are based on the method proposed in \cite{Luo:2019szz}, which in turns is based on decomposition of light-cone correlators into phase space integration of collinear splitting amplitudes of different multiplicities. The advantage of this decomposition is that it allows better understanding of the analytic continuation property of TMD PDFs and FFs~\cite{Chen:2020uvt}. Using the analytic continuation prescription of \cite{Chen:2020uvt}, we successfully obtained the N$^3$LO TMD FFs from corresponding TMD PDFs, without the need to compute everything from scratch. Our results open the avenue for precision phenomenology of TMD physics at N$^3$LO in perturbative QCD. 

We also provide threshold and high energy asymptotics of TMD PDFs and FFs through N$^3$LO by expanding the corresponding analytic expressions. The high energy~(small-$z$) limit of TMD FFs features double logarithmic enhancement, in contrast to the single logarithmic enhancement of TMD PDFs. We resum the small-$z$ logarithms through NNLL accuracy using a method proposed in \cite{Vogt:2011jv}. The resummation leads to better behaved perturbative convergence for $z< 10^{-2}$.

Our method of calculation is general and is not limited to unpolarized distribution. Works towards to polarized TMD distributions at N$^3$LO are in progress.

\noindent \textbf{Note added}: A few days after this paper appeared, an independent calculation for unpolarized quark and gluon TMD FFs at N$^3$LO was submitted to arXiv~\cite{Ebert:2020qef}. During private communication, the Authors of \cite{Ebert:2020qef} uncovered a minor error in one of their routine for analytic continuation. After fixing it, they found full agreement with our results. 

\acknowledgments

We thank Duff Neill and Simone Marzani for useful discussion. We also thank Markus Ebert, Bernhard Mistlberger, Gherardo Vita for useful correspondence. 
This work was supported in part by the National Science Foundation of China under contract No.~11975200 and No.~11935013. T.Z.Y. also want to acknowledge the support from the Swiss National Science Foundation (SNF) under contract 200020-175595.

\appendix

\section{QCD Beta Function}
\label{sec:beta}

The QCD beta function is defined as
\begin{equation}
\frac{d\alpha_s}{d\ln\mu} = \beta(\alpha_s) = -2\alpha_s \sum_{n=0}^\infty \left( \frac{\alpha_s}{4 \pi} \right)^{n+1} \, \beta_n \, ,
\end{equation}
with~\cite{Baikov:2016tgj}
\begin{align}
\beta_0 &= \frac{11}{3} C_A - \frac{4}{3} T_F N_f \, , \nn
\\
\beta_1 &= \frac{34}{3} C_A^2 - \frac{20}{3} C_A T_F N_f - 4 C_F T_F N_f \, ,\nn
\\
\beta_2 &= \left(\frac{158 C_A}{27}+\frac{44 C_F}{9}\right) N_f^2 T_F^2 +\left(-\frac{205 C_A
   C_F}{9}-\frac{1415 C_A^2}{27}+2 C_F^2\right) N_f T_F  +\frac{2857 C_A^3}{54}\,.
\end{align}

\section{Anomalous dimension}
\label{sec:AD}

For all the anomalous dimensions entering the renormalization group equations of various TMD functions, we define the perturbative expansion in $\alpha_s$ according to
\begin{equation}
\gamma(\alpha_s) = \sum_{n=0}^\infty \left( \frac{\alpha_s}{4 \pi} \right)^{n+1} \, \gamma_n \, ,
\end{equation}
where the coefficients for quark  are given by
\begin{align}
\Gcusp_{0} =& 4 C_F\,, \nn
\\
\Gcusp_{1} =&  \left(\frac{268}{9}-8 
                 \zeta_2\right) C_A C_F -\frac{80 C_F T_F N_f}{9}\,, \nn
\\
\Gcusp_{2} =&\bigg[ \left(\frac{320 \zeta _2}{9}-\frac{224 \zeta _3}{3}-\frac{1672}{27}\right) C_A
   C_F+\left(64 \zeta _3-\frac{220}{3}\right) C_F^2\bigg] N_f T_F  \nn
   \\
 +&\left(-\frac{1072 \zeta
   _2}{9}+\frac{88 \zeta _3}{3}+88 \zeta _4+\frac{490}{3}\right) C_A^2 C_F  -\frac{64}{27} C_F
   N_f^2 T_F^2\,,\nn
     \\
\gamma^S_0 =& 0 \, , \nn
\\
\gamma^S_1 =& \left[ \left( -\frac{404}{27} + \frac{11\zeta_2}{3} + 14\zeta_3 \right) C_A   + \left( \frac{112}{27} - \frac{4 \zeta_2}{3} \right)T_F N_f   \right]  C_F   \,, \nn
\\
\gamma^S_2  =&\left(-\frac{88}{3} \zeta
   _3 \zeta _2+\frac{6325 \zeta _2}{81}+\frac{658 \zeta _3}{3}-88 \zeta
   _4-96 \zeta _5-\frac{136781}{1458}\right) C_A^2 C_F
+\left(\frac{80\zeta _2}{27}-\frac{224 \zeta _3}{27}\right.
\nn\\
+&\left.\frac{4160}{729}\right) C_FN_f^2 T_F^2\nn
    + \left(-\frac{2828 \zeta _2}{81}-\frac{728 \zeta _3}{27}+48 \zeta
   _4+\frac{11842}{729}\right) C_A C_F N_f T_F
   \nn\\
   +&\left(-4 \zeta _2-\frac{304 \zeta _3}{9}-16 \zeta
   _4+\frac{1711}{27}\right) C_F^2 N_f T_F\,.\nn
\nn\\
   \gamma^R_0 = &0 \, , \nn
\\
\gamma^R_1 = & \left[ \left( -\frac{404}{27} + 14\zeta_3 \right) C_A  +  \frac{112}{27} T_F N_f \right] C_F  \,, \nn 
\\
\gamma^R_2 =&\bigg[\left(-\frac{824 \zeta _2}{81}-\frac{904 \zeta _3}{27}+\frac{20 \zeta
   _4}{3}+\frac{62626}{729}\right) C_A N_f T_F 
   +\left(-\frac{88}{3} \zeta _3 \zeta
   _2+\frac{3196 \zeta _2}{81}+\frac{6164 \zeta _3}{27}\right.
   \nn\\
   +&\left.\frac{77 \zeta _4}{3}-96 \zeta_5
   -\frac{297029}{1458}\right)C_A^2 
   + \left(-\frac{304 \zeta _3}{9}-16 \zeta_4+\frac{1711}{27}\right)  C_F N_f T_F 
   +\left(-\frac{64 \zeta_3}{9}\right.
   \nn\\
   -&\left.\frac{3712}{729}\right) N_f^2 T_F^2 \bigg] C_F \,.
\end{align}
Since cusp and  soft and rapidity anomalous dimensions exhibit Casimir scaling,
 the corresponding anomalous dimensions for gluon could be obtained by multiplying in above with $C_A/C_F$.

The beam anomalous dimensions do not exhibits Casimir scaling, thus should be list separately.
The beam anomalous dimensions  for quark are
 \begin{align}
\gamma^B_0 =& 3C_F \, , \nn
\\
\gamma^B_1 =&  \left[  \left( \frac{3}{2} - 12\zeta_2 + 24\zeta_3 \right) C_F + \left( \frac{17}{6} + \frac{44 \zeta_2}{3} - 12\zeta_3 \right)  C_A 
 + \left( -\frac{2}{3} - \frac{16\zeta_2}{3} \right) T_F N_f  \right] C_F  , \nn
\\
\gamma^B_2 = &
 \bigg[ \left(-\frac{2672 \zeta _2}{27}+\frac{400 \zeta _3}{9}+4
   \zeta _4+40\right) C_A C_F+\left(\frac{40 \zeta _2}{3}-\frac{272 \zeta
   _3}{3}+\frac{232 \zeta _4}{3}-46\right) C_F^2\bigg] N_f T_F \nn 
\\   
  +& \left(16 \zeta _3
   \zeta _2-\frac{410 \zeta _2}{3}+\frac{844 \zeta _3}{3}-\frac{494 \zeta
   _4}{3}+120 \zeta _5+\frac{151}{4}\right) C_A C_F^2
   +\left(\frac{320 \zeta _2}{27}-\frac{64 \zeta _3}{9}-\frac{68}{9}\right)
   \nn\\
  \times&  C_F N_f^2T_F^2
  + \left(\frac{4496
   \zeta _2}{27}-\frac{1552 \zeta _3}{9}-5 \zeta _4+40 \zeta _5-\frac{1657}{36}\right) C_A^2 C_F
   +\bigg(-32 \zeta _3 \zeta _2+18 \zeta _2
   \nn\\
   +&68 \zeta _3+144 \zeta_4-240 \zeta _5+\frac{29}{2}\bigg) C_F^3\,.
\end{align}
The beam anomalous dimensions  for gluon are
 \begin{align}
\gamma^B_0 =& \frac{11}{3} C_A - \frac{4}{3} T_F N_f \,, \nn
\\
\gamma^B_1 =& C_A^2 \left( \frac{32}{3}+ 12 \zeta_3 \right) + \left(  -\frac{16}{3} C_A -  4 C_F \right) N_f T_F   \,, \nn
\\
\gamma^B_2 =& C_A^3\left(-80\zeta_5-16\zeta_3\zeta_2+\frac{55}{3}\zeta_4+\frac{536}{3}\zeta_3+\frac{8}{3}\zeta_2+\frac{79}{2}\right)
\nn\\
+&C_A^2 N_f T_F\left(-\frac{20}{3}\zeta_4-\frac{160}{3}\zeta_3-\frac{16}{3}\zeta_2-\frac{233}{9}\right)
+\frac{58}{9}C_A N_f^2 T_F^2
-\frac{241}{9}C_A C_F N_f T_F
\nn\\
+&2 C_F^2 N_f T_F +\frac{44}{9}C_F N_f^2 T_F^2\,.
\end{align}
The cusp anomalous dimension $\Gamma^{\text{cusp}}$ can be found in \cite{Moch:2004pa}. 
 The beam anomalous dimension $\gamma^B$ is related to the soft anomalous dimension $\gamma^S$~\cite{Li:2014afw}
  and  the hard anomalous dimensions $\gamma^H$~\cite{Moch:2005tm,Gehrmann:2010ue,Becher:2009qa} by renormalization group invariance condition $\gamma^B = \gamma^S - \gamma^H$.
 The rapidity anomalous dimension $\gamma^R$ can be found in \cite{Li:2016ctv,Vladimirov:2016dll}. 
 Note that the normalization here differ from those in \cite{Li:2016ctv} by a factor of $1/2$. 

\section{Renormalization Constants}
\label{sec:RC}
The following constants are needed for the renormalization of zero-bin subtracted~\cite{Manohar:2006nz} TMD PDFs through N$^3$LO, see e.g. Ref.~\cite{Luo:2019hmp,Luo:2019bmw}. 
The first three-order corrections to $Z^B $ and $Z^S$ are 
\begin{align}
\label{eqZqZs}
Z^B_1 =& \frac{1}{2\epsilon} \left(2 \gamma^B_0 -\Gamma_0^{\text{cusp}} L_Q \right) \,, \nonumber \\
Z^B_2 =& \frac{1}{8 \epsilon^2} \bigg( ( \Gamma_0^{\text{cusp}} L_Q - 2 \gamma^B_0)^2 + 2 \beta_0 (  \Gamma_0^{\text{cusp}} L_Q - 2 \gamma^B_0)    \bigg)
 + \frac{1}{4\epsilon} \left( 2 \gamma^B_1 - \Gamma_1^{\text{cusp}} L_Q \right) \,, \nonumber \\
Z^B_3  =& \frac{1}{48 \epsilon^3}  \left( 2 \gamma^B_0 -\Gamma^{\text{cusp}}_0 L_Q \right) \biggl( 8 \beta_0^2 + 6 \beta_0 \left( -2 \gamma^B_0 + \Gamma^{\text{cusp}}_0 L_Q \right) 
+ \left( -2 \gamma^B_0 + \Gamma^{\text{cusp}}_0 L_Q \right)^2 \biggl) \nn \\ 
+& \frac{1}{24 \epsilon^2} \biggl( \beta_1 \left(-8 \gamma^B_0 + 4  \Gamma^{\text{cusp}}_0 L_Q \right) + \left(4 \beta_0 - 6 \gamma^B_0 + 3 \Gamma^{\text{cusp}}_0 L_Q \right) \left( -2 \gamma^B_1 + \Gamma^{\text{cusp}}_1 L_Q \right)  \biggl) 
 \nn  \\    
  +& \frac{1}{6 \epsilon} \biggl(  2 \gamma^B_2 -  \Gamma^{\text{cusp}}_2 L_Q  \biggl) 
  \nn\\  
Z^S_1 =& \frac{1}{\epsilon^2} \Gamma^{\text{cusp}}_0  +  \frac{1}{\epsilon} \left( -2 \gamma^S_0 - \Gamma_0^{\text{cusp}} L_\nu \right) \,,\nonumber \\
Z^S_2 =& \frac{1}{2 \epsilon^4} (\Gamma^{\text{cusp}}_0)^2 - \frac{1}{4 \epsilon^3} \bigg(\Gamma^{\text{cusp}}_0 (3 \beta_0 + 8 \gamma^S_0)+4( \Gamma^{\text{cusp}}_0)^2 L_\nu\bigg)   - \frac{1}{2 \epsilon} \left( 2 \gamma^S_1 +  \Gamma^{\text{cusp}}_1 L_\nu \right) \nonumber \\
+& \frac{1}{4 \epsilon^2} \bigg(\Gamma^{\text{cusp}}_1 + 2 ( 2 \gamma^S_0 + \Gamma^{\text{cusp}}_0 L_\nu ) ( \beta_0 + 2 \gamma^S_0 + \Gamma^{\text{cusp}}_0 L_\nu) \bigg) \,, \nn \\
Z^S_3 =& \frac{ 1}{6 \epsilon^6} \left(\Gamma^{\text{cusp}}_0\right)^3  - \frac{1}{4 \epsilon^5} \left(\Gamma^{\text{cusp}}_0 \right)^2 \left(  3 \beta_0 + 4 \gamma^S_0 + 2 \Gamma^{\text{cusp}}_0 L_\nu \right) + \frac{1}{36 \epsilon^4 } \Gamma^{\text{cusp}}_0 \bigg( 22 \beta_0^2 + 45 \beta_0 \left(2 \gamma^S_0 + \Gamma^{\text{cusp}}_0 L_\nu \right)  \nn \\
+& 9 \left( \Gamma^{\text{cusp}}_1 + 2 \left( 2 \gamma^S_0+ \Gamma^{\text{cusp}}_0 L_\nu\right)^2  \right)   \biggl)  + \frac{1}{36 \epsilon^3} \biggl( -16 \beta_1 \Gamma^{\text{cusp}}_0 - 12 \beta_0^2 \left( 2 \gamma^S_0 + \Gamma^{\text{cusp}}_0 L_\nu \right) \nn \\
 -& 2 \beta_0 \left( 5 \Gamma^{\text{cusp}}_1 + 9 \left( 2 \gamma^S_0 + \Gamma^{\text{cusp}}_0 L_\nu \right)^2 \right) - 3 \bigg[ \Gamma^{\text{cusp}}_1 \left(6 \gamma^S_0 + 9 \Gamma^{\text{cusp}}_0 L_\nu \right)  \nn \\
+& 2 \left( 8 \left( \gamma^S_0\right)^3 + 6 \Gamma^{\text{cusp}}_0 \gamma^S_1 + 12 \Gamma^{\text{cusp}}_0  \left(\gamma^S_0\right)^2 L_\nu + 6 \left(\Gamma^{\text{cusp}}_0\right)^2 \gamma^S_0 L_\nu^2 + \left( \Gamma^{\text{cusp}}_0 \right)^3 L_\nu^3 \right) \bigg]    \biggl) \nn \\
 +& \frac{1}{18 \epsilon^2} \biggl(  2 \Gamma^{\text{cusp}}_2 + 3 \left( 2 \beta_1 \left( 2 \gamma^S_0 + \Gamma^{\text{cusp}}_0 L_\nu \right) + \left( 2 \beta_0 + 6 \gamma^S_0 + 3 \Gamma^{\text{cusp}}_0 L_\nu \right) \left( 2 \gamma^S_1 + \Gamma^{\text{cusp}}_1 L_\nu \right) \right)    \biggl) 
 \nn\\
 - & \frac{2 \gamma^S_2 + \Gamma^{\text{cusp}}_2 L_\nu}{3 \epsilon} \,.
\end{align}
Keep in mind that the  anomalous dimensions appeared above depends on the flavor, they should be replaced by the corresponding values in Sec~\ref{sec:AD}.
We also remind the reader that the renormalization constants are formally identical for TMD PDFs and TMD FFs,
the logarithms appeared above should be replaced by their corresponding values in each case,
and we have
\begin{align}
\label{eq:LdefinitionALL}
 L_\perp = \ln \frac{b_T^2 \mu^2}{b_0^2} ,  \quad L_\nu = \ln \frac{\nu^2}{\mu^2} \,,
\end{align}
with $b_0 =2 \, e^{- \gamma_E}$ for both TMD PDFs and TMD FFs.

For  TMD PDFs,
\begin{align}
\label{eq:LdefinitionSLQ}
 L_Q  = 2 \ln \frac{x \, P_+}{\nu} \,,
\end{align}
while for TMD FFs,
\begin{align}
\label{eq:LdefinitionTLQ}
L_Q = 2 \ln \frac{P_+}{ z \, \nu}.
\end{align}

\bibliographystyle{JHEP}
\bibliography{unpolarizedN3LO}

\end{document}